\documentclass[twocolumn]{aastex63}
\usepackage{color}
\usepackage[titletoc]{appendix}
\usepackage[fleqn]{amsmath}
\usepackage{amssymb}
\usepackage{mathtools}
\usepackage{upgreek}
\usepackage{float}
\usepackage{comment}
\usepackage{enumitem}
\usepackage{natbib}
\usepackage{graphicx}
\usepackage{bm}
\usepackage{totcount}

\newtotcounter{citnum} %From the package documentation
\def\oldbibitem{} \let\oldbibitem=\bibitem
\def\bibitem{\stepcounter{citnum}\oldbibitem}

\shortauthors{Millholland \& Spalding}
\shorttitle{USPs from Obliquity Tides}

\begin{document} 

\title{Formation of Ultra-Short-Period Planets by Obliquity-Driven Tidal Runaway} 
\author{Sarah C. Millholland}
\altaffiliation{NASA Sagan Fellow}
\affiliation{Department of Astrophysical Sciences, Princeton University, Princeton, NJ 08544, USA}
\affiliation{Department of Astronomy, Yale University, New Haven, CT 06511, USA}

\author{Christopher Spalding}
\altaffiliation{51 Pegasi b Fellow}
\affiliation{Department of Astrophysical Sciences, Princeton University, Princeton, NJ 08544, USA}
\affiliation{Department of Astronomy, Yale University, New Haven, CT 06511, USA}
\email{sarah.millholland@princeton.edu}

\begin{abstract}
Small, rocky planets have been found orbiting in extreme proximity to their host stars, sometimes down to only $\sim 2$ stellar radii. These ultra-short-period planets (USPs) likely did not form in their present-day orbits, but rather migrated from larger initial separations. While tides are the probable cause of this migration, the tidal source has remained uncertain. Here we introduce planetary obliquity tides as a natural pathway for the production of USPs within close-in multi-planet systems. The crucial idea is that tidal dissipation generally forces planetary spin vectors to equilibrium configurations called ``Cassini states'', in which the planetary obliquities (axial tilts) are non-zero. 
%A significantly non-zero obliquity then leads to sustained tidal dissipation and orbital migration. 
In these cases, sustained tidal dissipation and inward orbital migration are inevitable. Migration then increases the obliquity and strengthens the tides, creating a positive feedback loop.
%leads to continued obliquity excitation and even stronger tides. 
Thus, if a planet’s initial semi-major axis is small enough ($a \lesssim 0.05$ AU), it can experience runaway orbital decay, which is stalled at ultra-short orbital periods when the forced obliquity reaches very high values ($\sim 85^{\circ}$) and becomes unstable. We use secular dynamics to outline the parameter space in which the innermost member of a prototypical \textit{Kepler} multiple-planet system can become a USP. We find that these conditions are consistent with many observed features of USPs, such as period ratios, mutual inclinations, and occurrence rate trends with stellar type. Future detections of stellar obliquities and close-in companions, together with theoretical explorations of the potential for chaotic obliquity dynamics, can help constrain the prevalence of this mechanism. \\

\end{abstract}

\section{Introduction}
\label{sec: Introduction}

%The discovery of small planets with periods  is one of many mysterious produced by the last decade of exoplanetary observations.
Small planets with extremely short orbital periods form a rare and fundamentally mysterious class of exoplanets. Notable early examples include CoRoT-7 b \citep{2009A&A...506..287L}, 
which at its discovery had the smallest radius ($1.7 \ R_{\oplus}$) and shortest orbital period ($0.85$ days) of any known planet, 
55 Cancri e \citep{2010ApJ...722..937D}, and Kepler-10 b \citep{2011ApJ...729...27B}. These ``ultra-short period planets'' (USPs), typically defined simply as planets with orbital periods $P < 1$ day \citep{2018NewAR..83...37W}, are about as rare as hot Jupiters. Unlike hot Jupiters, however, their occurrence rate increases for smaller stars; they exist around approximately $0.51\pm0.07\%$ of G-dwarf stars and $0.83\pm0.18\%$ of K-dwarf stars \citep{2014ApJ...787...47S}. Moreover, USP host stars do not exhibit the enhanced metallicity trend seen in hot Jupiter hosts \citep{2017AJ....154...60W}, and they almost always host additional planets %(albeit not always co-transiting planets) 
within $P<50$ days \citep{2014ApJ...787...47S}.

USPs are not likely to have formed where they're found. Their present-day orbits lie interior to the dust sublimation radius of typical protoplanetary disks, suggesting that these objects assembled on wider orbits before undergoing inward migration. Additional observational evidence supports this interpretation. When found in systems of multiple transiting planets, the period ratio between a USP and its nearest neighbor is usually $P_2/P_1 \gtrsim 4$ \citep{2013ApJ...774L..12S, 2018NewAR..83...37W}, larger than the $P_{j+1}/P_j\sim1.3-4$ typically seen in \textit{Kepler} systems of short-period planets with $P\sim1-100$ days \citep{2014ApJ...790..146F}, which are the USPs' closest counterparts. 

USPs are statistically distinct from the \textit{Kepler} multis in several additional respects, providing further evidence that they experienced a fundamentally different evolutionary history \citep{2016PNAS..11312023S}. The period distribution for planets with $P\lesssim1$ day follows a steeper power law \citep{2014ApJ...787...47S, 2017ApJ...842...40L, 2019MNRAS.488.3568P} compared to that at $P\sim1-10$ days, which is itself significantly different than the power law at $P\sim10-100$ days \citep{2018AJ....155...89P}. The USP radius distribution is also notable in that USPs are largely super-Earths, not sub-Neptunes; that is, the planets are almost always smaller than $R_p \lesssim 1.8 \ R_{\oplus}$ \citep{2014ApJ...787...47S, 2016NatCo...711201L}, on the smaller end of the observed radius valley \citep{2017AJ....154..109F}. This is evidence that any initial envelope of hydrogen/helium was lost either through photoevaporation driven by high-energy stellar irradiation \citep[e.g.][]{2017ApJ...847...29O, 2017MNRAS.472..245L} and/or heat from formation \citep[e.g.][]{2018MNRAS.476..759G}, leaving USPs as bare, rocky cores. These cores are observationally consistent with predominantly Earth-like compositions \citep{2019ApJ...883...79D}. 

Finally, USPs have larger mutual inclinations than more distant \textit{Kepler} multis \citep{2018ApJ...864L..38D}. This has been attributed to the gravitational influence of the stellar quadrupolar potential \citep{2020ApJ...890L..31L}, which is strong early on ($\lesssim 1$ Gyr) when the star is highly oblate due to its rapid rotation \citep{2016ApJ...830....5S, 2020AJ....160..105S}. This explanation requires that USPs reach their current orbits within $\sim$1 Gyr, thus favoring a fast migration process over a slow one. %Using Galactic velocity dispersions as a statistical proxy for stellar age, 
Additionally, \cite{2020arXiv200710944H} found that the ages of USP stellar hosts are indistinguishable from field star ages, which again supports an origin scenario faster than $\sim1$~Gyr.

Most proposed origins of the present-day orbits of USPs involve inward migration driven by tidal dissipation.\footnote{Several alternative USP origin theories have been proposed over the years, including that USPs are the remnant cores of hot Jupiters that underwent Roche lobe overflow \citep{2013ApJ...779..165J, 2016CeMDA.126..227J,  2014ApJ...793L...3V, 2017ApJ...846L..13K}. This is now disfavored based on the lack of correlation between USP occurrence and stellar metallicity \citep{2017AJ....154...60W}.} The primary source of this dissipation remains unclear. One proposed source is stellar tides. In particular, \cite{2017ApJ...842...40L} posited that the proto-USP planets could form in situ near the innermost edge of the protoplanetary disk 
%(which is likely controlled by magnetospheric truncation)
and migrate inwards due to tides raised in the star. With stellar tides alone, however, generating USPs from initial orbits $P>1$ days would require stellar quality factors that are inconsistent with observational estimates \citep{2010ApJ...723..285H, 2012ApJ...751...96P, 2019AJ....157..180P, 2019MNRAS.488.3568P}. Moreover, \cite{2020arXiv200710944H}'s finding that USP hosts have similar ages as field stars implies that USPs are generally stable against inspiral from stellar tides.

Another set of theories have explored tidal dissipation raised in the planet (i.e. planetary tides) as opposed to the star, which is stronger for planets in the USP mass regime. After the detection of CoRoT-7 b, but before many other USPs had been found, \cite{2010ApJ...724L..53S} proposed that dynamical interactions in multi-planet systems could scatter super-Earths to short-period and eccentric orbits, at which point tides would lead to further orbital decay and circularization. More recently, \cite{2019AJ....157..180P} proposed that USPs form in multi-planet systems with initial periods of $\sim5-10$ days, before undergoing chaotic secular interactions that cause them to reach high eccentricities. Strong tidal dissipation then induces high eccentricity migration, ending with the planets on roughly circular orbits at very short periods. \cite{2019MNRAS.488.3568P} examined a similar scenario of eccentricity-based tidal migration driven by secular planet-planet interactions, but they suggested a dynamically cooler evolution with eccentricities $e\sim0.1-0.4$ and initial $\sim 1-3$ day orbital periods. 

The present-day eccentricities of \textit{Kepler} close-in, multi-transiting systems are generally quite low, $\bar{e}\sim0.04$  \citep{2016PNAS..11311431X, 2019AJ....157...61V, 2019AJ....157..198M}. Though their primordial values may have differed, stability arguments suggest similar values in order to match the observed system architectures \citep{2019MNRAS.484.1538W}. It is thus worthwhile to consider a USP formation scenario that could operate without any requirement on eccentricities. Moreover, this would avoid the complications of tidal disruption, which can be problematic for the high eccentricity migration scenario \citep{owen2018photoevaporation}, and orbital instability, which is a risk for the short-period, tightly-packed systems that USPs are often found in \citep[e.g.][]{2016AJ....152..105M}. In particular, observed USPs in multi-transiting systems often have companions with $P<10$ days \citep{2018NewAR..83...37W}, which our new theory of USP production will aim to account for.

Apart from stellar tides and planetary eccentricity tides, a source of tidal dissipation that has not yet been considered is planetary obliquity tides. Here, the ``planetary obliquity’’ refers to the axial tilt of the planet’s spin axis off its orbital axis ($\sim23^{\circ}$ for Earth).\footnote{Throughout this work, we will refer to the planetary obliquity simply as the ``obliquity''. We will use the term ``stellar obliquity'' when referencing the angle between the stellar spin and the orbital axes.} Both eccentricity and obliquity tides are important components of the overall tidal dissipation rate \citep[e.g.][]{2005ApJ...628L.159W, 2008Icar..193..637W, 2010A&A...516A..64L, 2019NatAs...3..424M}. However, a critical feature is that, unlike eccentricity tides, the equilibrium state of obliquity tides is not generally a zero obliquity.  

%A crucial insight is that, in multi-planet system configurations with non-zero mutual inclinations, planetary obliquities are \textit{forced} to be non-zero and obliquity tides are therefore \textit{inevitable}.
In short-period, multi-planet systems with non-zero mutual orbital inclinations, tidal dissipation leads planetary obliquities to non-zero states, making continued dissipation via obliquity tides inevitable \citep[e.g.][]{1974AJ.....79..722P}. This arises as a consequence of orbital precession induced by secular interactions. In an inclined and precessing orbit frame, the equilibrium positions of a planet’s spin vector have non-zero obliquities. Often the forced obliquities are $\lesssim 1^{\circ}$, but sometimes they are much larger (e.g. $\gtrsim 10^{\circ}$), particularly if the mutual orbital inclinations are large.  These equilibrium configurations of the spin vector are called ``Cassini states'' \citep{1969AJ.....74..483P}, and tidal dissipation will rapidly force short-period planets to occupy them. Historically, Cassini states were first studied in the context of the Moon \citep{1966AJ.....71..891C} and thereafter in other Solar System bodies \citep[e.g.][]{1969AJ.....74..483P, 1974AJ.....79..722P, 1975AJ.....80...64W}. Recent works have explored Cassini states 
% has been invoked as in the context of
within short-period exoplanetary systems and shown that these forced non-zero obliquities could help explain several disparate mysteries \citep{2018ApJ...869L..15M, 2019NatAs...3..424M, 2019ApJ...876..119M}.

Most often, non-zero obliquities do not affect orbital evolution substantially. However, for short-period planets, obliquities can manifest through sustained planetary tidal dissipation, generating semi-major axis decay and interior heating \citep{2019NatAs...3..424M, 2019ApJ...886...72M, 2020ApJ...897....7M}. If a planet begins in a $P\sim 1-5$ day orbit, obliquity tides can lead to rapid runaway orbital decay, a scenario recently proposed for the hot Jupiter WASP-12 b \citep{2018ApJ...869L..15M}. As the orbit shrinks, a high obliquity Cassini state evolves to even larger obliquities and eventually becomes unstable to tides \citep{2007ApJ...665..754F, 2008ASPC..398..281P}. The obliquity then damps back down to a separate Cassini equilibrium with a low (but non-zero) obliquity, thereby stalling the rapid orbital decay. 

In this paper, we show how obliquity tides driven by Cassini states with forced non-zero obliquities can naturally lead to rapid tidal migration of planets initially on $P\sim 1-5$ day orbits, turning them into USPs. This mechanism can act either as an accompaniment or alternative to eccentricity-based tidal migration. The paper is organized as follows. We begin by describing Cassini states and obliquity tides, before outlining their role in USP production (Section \ref{sec: Cassini states and obliquity tides}). We then use this theory to map out the parameter space in which the innermost member of a close-in, multi-planet system is susceptible to becoming a USP through obliquity tides (Section \ref{sec: Parameter Space for USP Production}). We examine USP planets in observed systems in Section \ref{sec: Observed USP Planets in Multi-transiting Systems} and discuss limitations of the theory in Section \ref{sec: Limitation: angular momentum budget}. We discuss observational predictions and further extensions, such as chaos and early system evolution, in Section \ref{sec: Discussion} and conclude in Section \ref{sec: Conclusion}.

\section{Cassini States and Obliquity Tides}
\label{sec: Cassini states and obliquity tides}

Our proposed mechanism of USP production via \textit{obliquity-driven tidal migration} can be divided into three stages, roughly representing the start, middle, and end: 
\begin{enumerate}[nolistsep, labelindent=0pt]
\item Initial entry into Cassini states
\item Tidal migration and evolution of a forced Cassini state obliquity 
\item Tidal breaking of Cassini states and stalling of migration at ultra-short period orbits 
\end{enumerate}
The following three sub-sections describe these stages. 

\subsection{Entry into Cassini states}
\label{sec: Entry into Cassini states}

The spin vectors of close-in planets are subject to a dissipative tidal torque that moves them towards equilibrium configurations. The tidal torque arises due to the gravitational deformation (or ``bulge'') raised on the planet from its host star; it is dissipative because it involves this bulge sweeping across the planet every orbit. If the orbit is static, the tidally-relaxed equilibrium of the spin vector is a straightforward spin-synchronous and aligned state, where the spin rotation frequency, $\omega = 2\pi/P_{\mathrm{rot}}$, is equal to the orbital mean motion, $n = 2\pi/P$, and the obliquity, $\epsilon$, is zero. However, most planetary orbits are not static; they undergo precession due to interactions with other planets, the oblate host star, and any other gravitational sources that cause deviations from a $1/r$ potential. The equilibrium configurations of the spin pole in a uniformly precessing orbit frame are known as ``Cassini states''. The dynamical origin and behavior of the Cassini states have been documented in many previous works \citep[e.g.][]{1966AJ.....71..891C, 1969AJ.....74..483P, 1974AJ.....79..722P, 1975AJ.....80...64W, 2004AJ....128.2501W, 2015A&A...582A..69C, 2020arXiv200414380S}. Here, we summarize the most relevant material.\footnote{We have emphasized here that Cassini states are reached as the end-product of tidal dissipation, but it is important to note that planets/satellites can also enter Cassini states through resonant capture and excitation.
%whenever the $|g|$ and $\alpha$ frequencies evolve and become commensurable. This resonant sweeping occurs through slow changes in $a$, $\omega$, $R_p$, $g$, or other system parameters. 
For example, this process is what is thought to have generated Saturn's $27^{\circ}$ obliquity \citep{2004AJ....128.2501W, 2004AJ....128.2510H}. The resonance is typically called a ``secular spin-orbit resonance'' in the literature \citep[e.g.][]{1993Sci...259.1294T, 2004AJ....128.2501W, 2019A&A...623A...4S, 2019NatAs...3..424M, 2019ApJ...876..119M}, and it is an instance of a Cassini state. In this work, we primarily use the term ``Cassini state'' rather than ``secular spin-orbit resonance'' so as to highlight that no resonant sweeping of frequencies is required to produce the Cassini state here.}

Cassini states are configurations in which the precession rate of the planet's spin axis exactly matches that of its orbital plane. More specifically, the planetary spin axis, $\bm{\hat{s}}$, and unit orbit normal vector, $\bm{\hat{n}}$, precess at the same rate about the axis of the total system angular momentum vector, $\bm{\hat{k}}$.  In a dissipationless Cassini state, these three vectors are coplanar. Dissipation causes $\bm{\hat{s}}$ to shift out of the plane defined by $\bm{\hat{n}}$ and $\bm{\hat{k}}$. For a given orbital inclination, $I$, with respect to the invariable plane, the obliquities of Cassini states obey the relation \citep[e.g.][]{1975AJ.....80...64W}
\begin{equation}
g\sin(\epsilon-I)+\alpha\cos\epsilon\sin\epsilon=0.
\label{eq: Cassini state relation}
\end{equation}
Here, $g = \dot{\Omega}$ is the precession frequency of the longitude of the ascending node. This frequency is negative (corresponding to nodal recession) for the cases of interest here. The frequency $\alpha$ is the spin-axis precession constant, which sets the precession period, $T_{\alpha}~=~2\pi/(\alpha\cos\epsilon)$, of the spin-axis due to the torque induced by the host star on the oblate planet. In the absence of satellites orbiting the planet, $\alpha$ is given by 
\citep{1997A&A...318..975N} 
\begin{equation}
\alpha=\frac{1}{2}\frac{M_{\star}}{M_p}\left(\frac{R_p}{a}\right)^3\frac{k_2}{C}\frac{\omega}{(1-e^2)^{3/2}}.
\label{eq: alpha}
\end{equation}
Here, $M_{\star}$ is the stellar mass, $M_p$ the planet mass, $R_p$ the planet radius, $a$ the semi-major axis, $e$ the eccentricity, $k_2$ the planetary Love number, and $C$ the planet's moment of inertia normalized by $M_p {R_p}^2$.

It is important to note that Cassini states are strictly only defined for uniform orbital precession, that is, when $g=\dot{\Omega}$ and $I$ are constant. When the precession is non-uniform, the planet's orbital inclination/node solution is composed of a superposition of several modes with multiple frequencies $\{g_i\}$ and amplitudes $\{I_i\}$. The resulting spin vector equilibria are ``quasi-Cassini states'', which behave approximately like Cassini states with $g$ in equation \ref{eq: Cassini state relation} equal to one of the $g_i$ modes. For example, Saturn's proposed Cassini state is associated with the $g_8$ inclination/node fundamental frequency of the Solar System, which is dominated by Neptune's nodal precession \citep{2004AJ....128.2501W, 2004AJ....128.2510H}. This multiple modes concept will be revisited several times in this work. For now, we will simply assume that the frequency $g$ corresponds to one of the ${g_i}$ secular modes of the system. 
\begin{figure}
\epsscale{1.}
\plotone{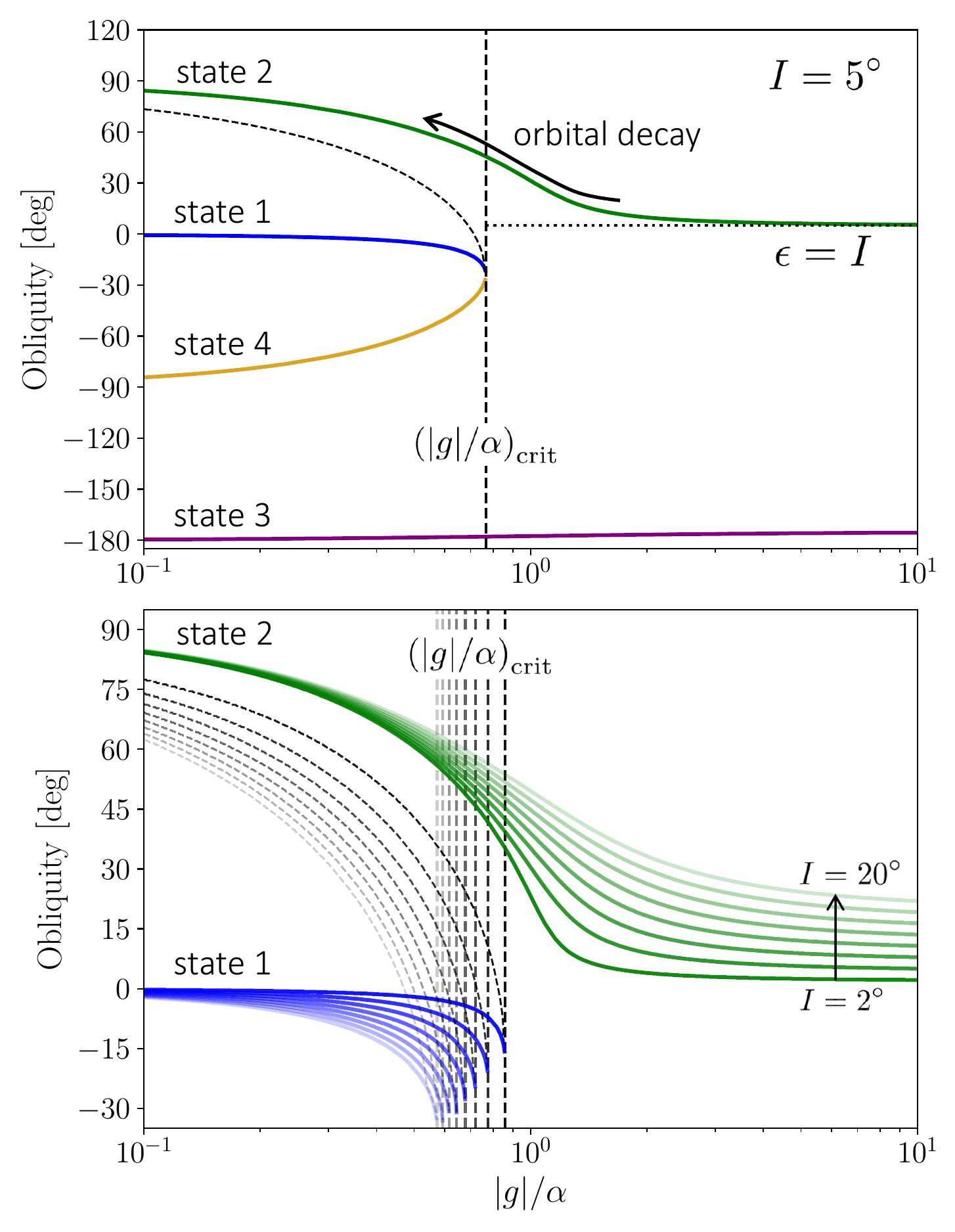}
\caption{The tidal equilibrium positions of the planetary obliquity as a function of $|g|/{\alpha}$. \textit{Top panel:} The obliquities of the four Cassini states are plotted versus $|g|/{\alpha}$ for an inclination of $I=5^{\circ}$. For $|g|/{\alpha} > \left(|g|/{\alpha}\right)_{\mathrm{crit}}$ (equation \ref{eq: g/alpha_crit}, vertical dashed line), states 1 and 4 disappear, and the obliquity of state 2 tends towards $\epsilon = I$ (horizontal dotted line). \textit{Bottom panel:} The variation of Cassini states 1 and 2 is shown as a function of inclination, using equal steps from $I=2^{\circ}$ (most opaque) to $I=20^{\circ}$ (most transparent). The key observations are (1) the absolute value of the obliquity of both states increases with $I$, particularly beyond $\left(|g|/{\alpha}\right)_{\mathrm{crit}}$; (2) the critical ratio decreases with $I$; and (3) orbital decay leads to a decrease in $|g|/{\alpha}$ and an increase in the Cassini state 2 obliquity.} 
\label{fig: Four_cassini_states}
\end{figure}

With $g$ and $\alpha$ frequencies specified, there are either two or four well-defined Cassini states, depending on the frequency ratio $|g|/\alpha$ in reference to the critical frequency ratio,
\begin{equation}
\left(|g|/{\alpha}\right)_{\mathrm{crit}} = (\sin^{2/3}I + \cos^{2/3}I)^{-3/2}.
\label{eq: g/alpha_crit}
\end{equation}
This critical ratio decreases as a function of $I$ for $I < 45^{\circ}$. When $|g|/{\alpha} < \left(|g|/{\alpha}\right)_{\mathrm{crit}}$, equation \ref{eq: Cassini state relation} has four roots, corresponding to Cassini states 1-4. When $|g|/{\alpha} > (|g|/{\alpha})_{\mathrm{crit}}$, equation \ref{eq: Cassini state relation} has two roots, corresponding to Cassini states 2 and 3. States 1 and 2 are stable; state 3 is linearly stable but unstable to tidal evolution \citep{2007ApJ...665..754F}; and state 4 is unstable. Thus, Cassini states 1 and 2 will be our primary focus, as they are the only ones that are stable in the presence of tides.

Cassini state 1 corresponds to the configuration in which $\bm{\hat{s}}$ and $\bm{\hat{n}}$ are on the same side of $\bm{\hat{k}}$. In this case, the convention is for the obliquity to be defined as negative. In state 2, $\bm{\hat{s}}$ and $\bm{\hat{n}}$ are on opposite sides of $\bm{\hat{k}}$, and the obliquity is positive. In the limit $|g|/{\alpha} \ll (|g|/{\alpha})_{\mathrm{crit}}$, the state 1 and 2 equilibrium obliquities are 
\begin{equation}
\epsilon_1 \approx \tan^{-1}\left(\frac{\sin{I}}{1+\alpha/g}\right) ; \ \ \ 
\epsilon_2 \approx \cos^{-1}\left(\frac{-g\cos{I}}{\alpha}\right). 
\label{eq: Cassini state 1 and 2}
\end{equation}

Figure \ref{fig: Four_cassini_states} shows the obliquities of the Cassini states as a function of $|g|/{\alpha}$, through solving equation \ref{eq: Cassini state relation}. The top panel shows all four states for $I = 5^{\circ}$, and the bottom panel shows the evolution of states 1 and 2 and $(|g|/{\alpha})_{\mathrm{crit}}$ with respect to $I$. It is important to emphasize that both states 1 and 2 have non-zero obliquities whenever $I>0^{\circ}$, making a misaligned planetary spin axis \textit{unavoidable}. However, state 1's obliquity is very close to zero for small $I$ and for $|g|/{\alpha} \lesssim 0.1$. Meanwhile, state 2 allows for very large obliquities. The absolute value of the obliquity of both states increases with $I$, and in the limit $|g|/{\alpha} \gg (|g|/{\alpha})_{\mathrm{crit}}$, the state 2 obliquity asymptotes at $\epsilon_2 = I$. Depending on the initial conditions, tides will carry the spin vector to either state 1 or state 2. When $|g|/{\alpha} > (|g|/{\alpha})_{\mathrm{crit}}$, state 2 is required. Thus, a high obliquity state is often the only option.

\subsection{Obliquity-driven tidal migration (and evolution of Cassini states)}
\label{sec: Obliquity-driven tidal migration}

%When a planet is trapped in a Cassini state with a non-zero obliquity, it inevitably experiences dissipation due to the tides raised on the planet from its host star; the gravitational tidal deformation (or ``bulge'') sweeps across the planet each orbit, leading to a buildup of interior friction and heat.

Assume, for now, that a planet initially tidally relaxes into Cassini state 2. (Later we will show that this is often the case.) Once this happens, the same dissipative tidal torque that brought the planet into the Cassini state will continue to generate heat in the planetary interior due to the non-zero obliquity. This dissipation affects the orbit too. It generates orbital decay, since the thermal energy dissipated in the interior is balanced by the conversion of orbital energy. The orbital decay, in turn, leads to a decrease in the ratio $|g|/{\alpha}$ (assuming that the orbital precession is arising from planet-planet interactions) and an increase in the equilibrium obliquity of Cassini state 2. This is depicted in the upper panel of Figure \ref{fig: Four_cassini_states}.

While tidal dissipation and orbital decay necessarily result from non-zero obliquities, quantifying the magnitude of energy dissipation and orbital decay is non-trivial, and there are many available tidal models \citep[e.g.][]{2009CeMDA.104..257E, 2013CeMDA.116..109F, 2014A&A...571A..50C}. The simplest and most widely-used approach is equilibrium tide theory \citep[e.g.][]{1880RSPT..171..713D, 1966Icar....5..375G, 1979M&P....20..301M, 1981A&A....99..126H, 1998ApJ...499..853E}, which we adopt here using the viscous approach \citep{2007A&A...462L...5L, 2010A&A...516A..64L}. 

The basic assumptions are that the planet's gravitational response to the tidal forces constitutes a hydrostatic deformation, or tidal bulge, and this bulge lags the star's position with a constant time lag, $\Delta t$. The constant time offset is often parameterized in terms of the annual tidal quality factor, $Q$, which is related to $\Delta t$ through $Q = (\Delta t n)^{-1}$. $Q$ quantifies the efficiency of tidal damping, and it is combined with $k_2$ into the ``reduced tidal quality factor'', $Q' = 3Q/2k_2$. $Q$ is highly uncertain for individual planets but is known in an order-of-magnitude sense for different planetary archetypes. Rocky bodies in the Solar System have $Q\sim100$ \citep{1966Icar....5..375G, 1999ssd..book.....M}, while extrasolar super-Earths and sub-Neptunes have been found with $Q\sim10^3-10^5$ \citep{2017AJ....153...86M, 2018AJ....155..157P}, similar to the estimated values for Uranus \citep{1989Icar...78...63T} and Neptune \citep{2008Icar..193..267Z}. We assume a range of plausible planetary $Q$ values in this work. As for $k_2$, constraints come from both the Solar System bodies \citep{2016CeMDA.126..145L} and from theoretical models \citep{2011A&A...528A..18K, 2018A&A...615A..39K}, which we will use to inform our fiducial values in this work.

Within the viscous equilibrium tide model, the tidal luminosity --- or the rate at which orbital energy is converted into thermal energy --- is given by the expression \citep{2007A&A...462L...5L, 2010A&A...516A..64L}:
\begin{equation}
L_{\mathrm{tide}}(e,\epsilon) = 2 K\left[N_a(e) - \frac{N^2(e)}{\Omega(e)}\frac{2\cos^2\epsilon}{1+\cos^2\epsilon}\right]. 
\label{eq: full dissipation rate}
\end{equation}
Here, $N_a(e)$, $N(e)$, $\Omega(e)$ are functions of eccentricity given by
\begin{align}
N_a(e) &= \frac{1 + \frac{31}{2}e^2 + \frac{255}{8}e^4 + \frac{185}{16}e^6 + \frac{25}{64}e^8}{(1-e^2)^{\frac{15}{2}}} \label{eq: N_a(e)} \\
N(e) &= \frac{1 + \frac{15}{2}e^2 + \frac{45}{8}e^4 + \frac{5}{16}e^6}{(1-e^2)^6} \label{eq: N(e)} \\
\Omega(e) &= \frac{1 + 3e^2 + \frac{3}{8}e^4}{(1-e^2)^{\frac{9}{2}}}.
\label{eq: Omega(e)}
\end{align}
The magnitude of $L_{\mathrm{tide}}$ is dictated by $K$, the characteristic luminosity scale,
\begin{equation}
K = \frac{3n}{2}\frac{k_2}{Q}\left(\frac{G {M_{\star}}^2}{R_p}\right)\left(\frac{R_p}{a}\right)^6. 
\label{eq: tidalK}
\end{equation}
Equation \ref{eq: full dissipation rate} assumes that the planet's spin rotation frequency has reached its equilibrium rate given by
\begin{equation}
\omega_{\mathrm{eq}}=n\frac{N(e)}{\Omega(e)}\frac{2\cos\epsilon}{1+\cos^2\epsilon}. 
\label{eq: omega_eq}
\end{equation}

The tidal luminosity is balanced by a decrease in the orbital energy, such that $L_{\mathrm{tide}}~=~-(GM_{\star}M_p\dot{a})/(2a^2)$. Using this we may calculate the orbital decay timescale,
\begin{equation}
\tau_a = \frac{a}{\dot{a}} = -\frac{G M_{\star}M_p}{4aK}\Big[N_a(e) - \frac{N^2(e)}{\Omega(e)}\frac{2\cos^2\epsilon}{1+\cos^2\epsilon}\Big]^{-1}.
\label{eq: tau_a}
\end{equation}

\begin{figure}
\epsscale{1.}
\plotone{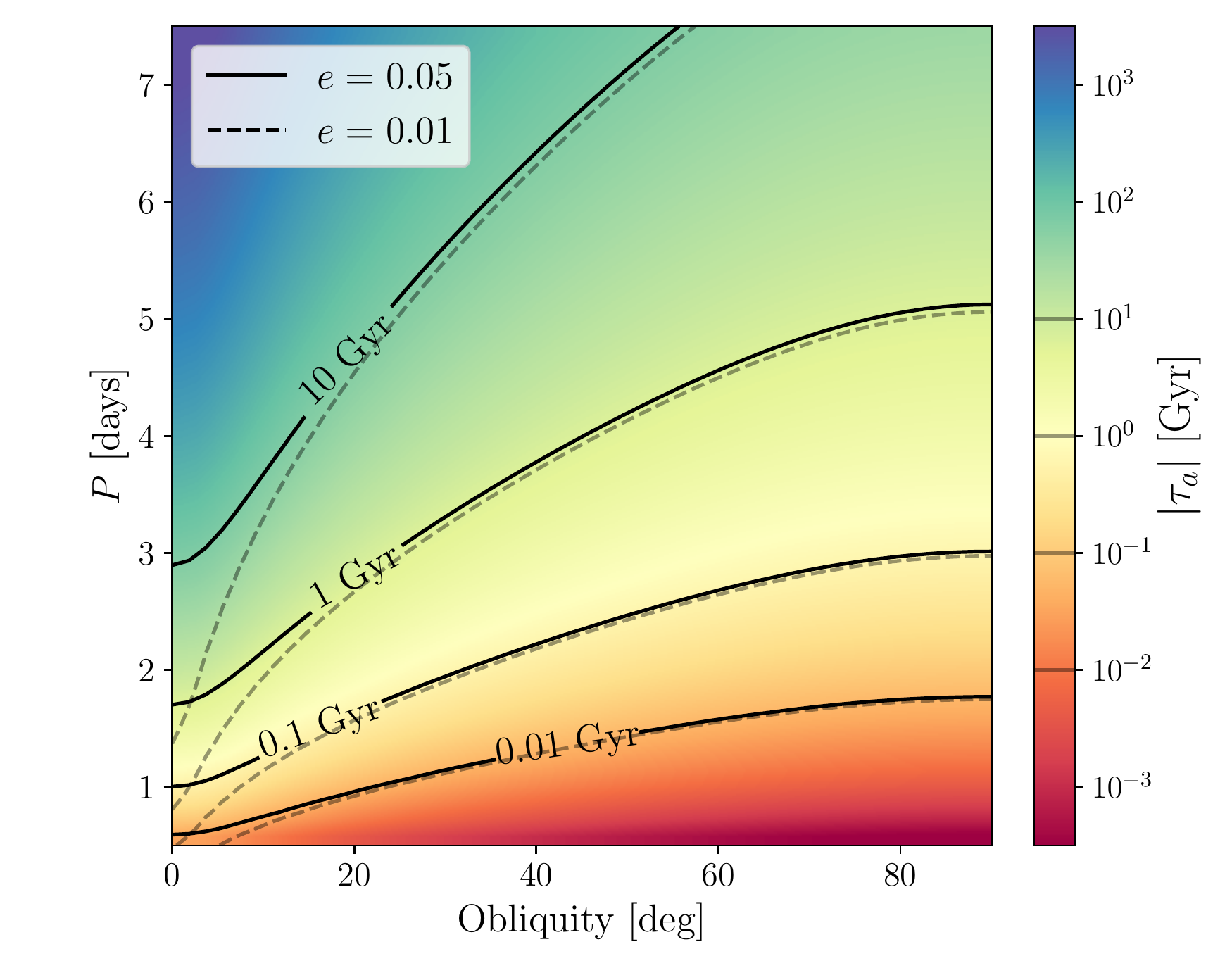}
\caption{Tidal migration timescale, $|\tau_a|$ (equation \ref{eq: tau_a}), as a function of the orbital period and planetary obliquity for typical parameters. The colormap corresponds to an eccentricity of $e=0.05$, which is also represented by the solid contour lines. We also show the $e=0.01$ case with dashed contour lines. This figure assumes the following set of fiducial parameters : $M_{\star} = M_{\odot}$, $M_p = 6 \ M_{\oplus}$, $R_p = 1.63 \ R_{\oplus}$ (calculated from $M_p$ using \citealt{2016ApJ...819..127Z}'s Earth-like planetary composition curve), $Q = 10^3$, and $k_2 = 0.4$. For sufficiently small $P$ and large enough $e$ and/or $\epsilon$, $|\tau_a|$ is fast enough to induce substantial migration during the system lifetime.} 
\label{fig: tau_a_heatmap}
\end{figure}

The $\tau_a$ timescale can be used to delineate the regime in which planets undergo significant tidal migration for a given initial semi-major axis. Note that all planets in the system, not just the innermost one, can be migrating due to non-zero obliquities and/or eccentricities. However, it is generally only the innermost planet that can migrate fast enough to substantially separate itself from its neighbors within the age $\tau_{\textrm{age}}$ of the system, since the $\tau_a$ timescale depends strongly on $a$. Figure \ref{fig: tau_a_heatmap} shows $|\tau_a|$ as a function of $P$ and $\epsilon$ for a fiducial set of rocky planet parameters. When $|\tau_a| \lesssim \tau_{\mathrm{age}} \approx 1 - 10$ Gyr, corresponding to $P\lesssim 2-3$ days for $\epsilon\approx10^{\circ}$,  the semi-major axis will decrease by order unity during the system lifetime. The instantaneous $|\tau_a|$ timescale is an overestimate of the total time to decay, however, since orbital migration is a runaway process in which $|\tau_a|$ decreases as the orbit shrinks and the obliquity (in Cassini state 2) increases. Accordingly, it is useful to calculate the time, $\tau_{\mathrm{decay}}$, for complete decay from some initial $a=a_i$ to an ending position with $a\approx0$. Using the fact that $\dot{a} \propto a^{-11/2}$ for constant obliquity and eccentricity, one can show that
\begin{equation}
\tau_{\mathrm{decay}} \approx \frac{2}{13}|\tau_a(a=a_i)|.
\label{eq: tau_decay approx}
\end{equation}
Thus, for some initial $a_i$, the timescale for complete decay of order $\Delta a \sim a_i$ is nearly an order of magnitude smaller than the decay timescale at the initial separation, $|\tau_a(a=a_i)|$. In practice, a more accurate estimate of $\tau_{\mathrm{decay}}$ than that provided by equation \ref{eq: tau_decay approx} can be obtained through numerical integration of $\dot{a}$ in equation \ref{eq: tau_a} using an evolving $e$ and $\epsilon$. This will be our approach in Section \ref{sec: Identifying the susceptible parameter space for USP production}.

An orbit will only decay completely, however, if it maintains a non-zero eccentricity and/or obliquity. If these go to zero, the tidal dissipation and migration will stall. This stalling is expected due to the tidal breaking of Cassini state 2 at high obliquity.

\subsection{Tidal breaking of Cassini state 2}
\label{sec: Tidal breaking of Cassini states}

Cassini state 2 cannot exist at an arbitrarily high obliquity in the presence of tides \citep{2007ApJ...665..754F, 2008ASPC..398..281P}. As the obliquity increases due to the orbital decay, there becomes a point at which the dissipative tidal torque overwhelms the orbital precession torque; Cassini state 2 breaks, and the obliquity damps down to Cassini state 1. \cite{2007ApJ...665..754F} showed that the breaking obliquity is related to the limits of a specific phase shift that appears in Cassini state 2 in the presence of steady tidal dissipation. The spin axis, $\bm{\hat{s}}$, shifts out of the plane containing $\bm{\hat{n}}$ and $\bm{\hat{k}}$, and this phase shift provides a balance of the tidal torque, up until the breaking point.

In order to identify this breaking obliquity, we consider the secular equation of motion of the spin vector, $\bm{\omega} = \omega\bm{\hat{s}}$, using the framework of \cite{2001ApJ...562.1012E} and \cite{2007ApJ...665..754F} with zero orbital eccentricity. The equation may be written as the sum of two torques: a non-dissipative torque due to the star's non-uniform gravitational force on the planet, and a dissipative tidal torque due to the lagged response of the tidal bulge raised in the planet. The non-dissipative torque generates the spin axis precession, and the dissipative tidal torque drives the spin vector towards the Cassini states. Explicitly, we may write
\begin{equation}
\dot{\bm{\omega}} = \dot{\bm{\omega}}_{\mathrm{prec}} + \dot{\bm{\omega}}_{\mathrm{tides}},
\label{eq: torque}
\end{equation}
where 
\begin{equation}
\begin{split}
\dot{\bm{\omega}}_{\mathrm{prec}} &= \alpha\omega(\bm{\hat{s}}\cdot\bm{\hat{n}})(\bm{\hat{s}} \times \bm{\hat{n}}) \\
\dot{\bm{\omega}}_{\mathrm{tides}} &= \frac{n}{C}\left(\frac{a}{R_p}\right)^2\left[-\frac{\bm{\omega}}{2 n t_F} + \left(1-\frac{\bm{\omega} \cdot \bm{\hat{n}}}{2n}\right)\frac{\bm{\hat{n}}}{t_F}\right].
\label{eq: torques}
\end{split}
\end{equation}
Here, $t_F$ is the tidal friction timescale given by
\begin{equation}
t_F = \frac{4Q'}{9}\left(\frac{a}{R_p}\right)^5\frac{M_p}{M_{\star}}\frac{1}{n}.
\end{equation}

%Considering the prefactors of the precession torque and the dissipative torque is...  $|\dot{\bm{\omega}}_{\mathrm{prec}}|/|\dot{\bm{\omega}}_{\mathrm{tides}}| \approx Q$. That is, all dependence on other stellar or planetary parameters (such as $M_{\star}$, $R_p$, $a$, etc.) drop out of the ratio since $\dot{\bm{\omega}}_{\mathrm{prec}}$ and $\dot{\bm{\omega}}_{\mathrm{prec}}$ depend on them in the same way. However, the vector evolution is also a function of $I$, so the breaking obliquity depends on $Q$ and $I$ alone. 

Inspecting the expressions for $\dot{\bm{\omega}}_{\mathrm{prec}}$ and $\dot{\bm{\omega}}_{\mathrm{tides}}$, we observe that both torques exhibit the same dependencies with most physical parameters of the problem, including $M_{\star}$, $M_p$, $R_p$, $a$, $k_2$, and $C$. The two exceptions are $Q$, which enters into $\dot{\bm{\omega}}_{\mathrm{tides}}$ but not $\dot{\bm{\omega}}_{\mathrm{prec}}$, and $I$, which factors into the equations via $\bm{\hat{n}}$. These dependencies imply that the breaking obliquity depends only on $Q$ and $I$. Accordingly, in order to identify the breaking obliquity, we can simply select arbitrary system parameters and numerically integrate equation~\ref{eq: torque} in response to an evolving $|g|/{\alpha}$. Doing this for different values of $Q$ and $I$ and determining the maximum obliquity in each case will provide the full range of outcomes.

\begin{figure}
\epsscale{1.}
\plotone{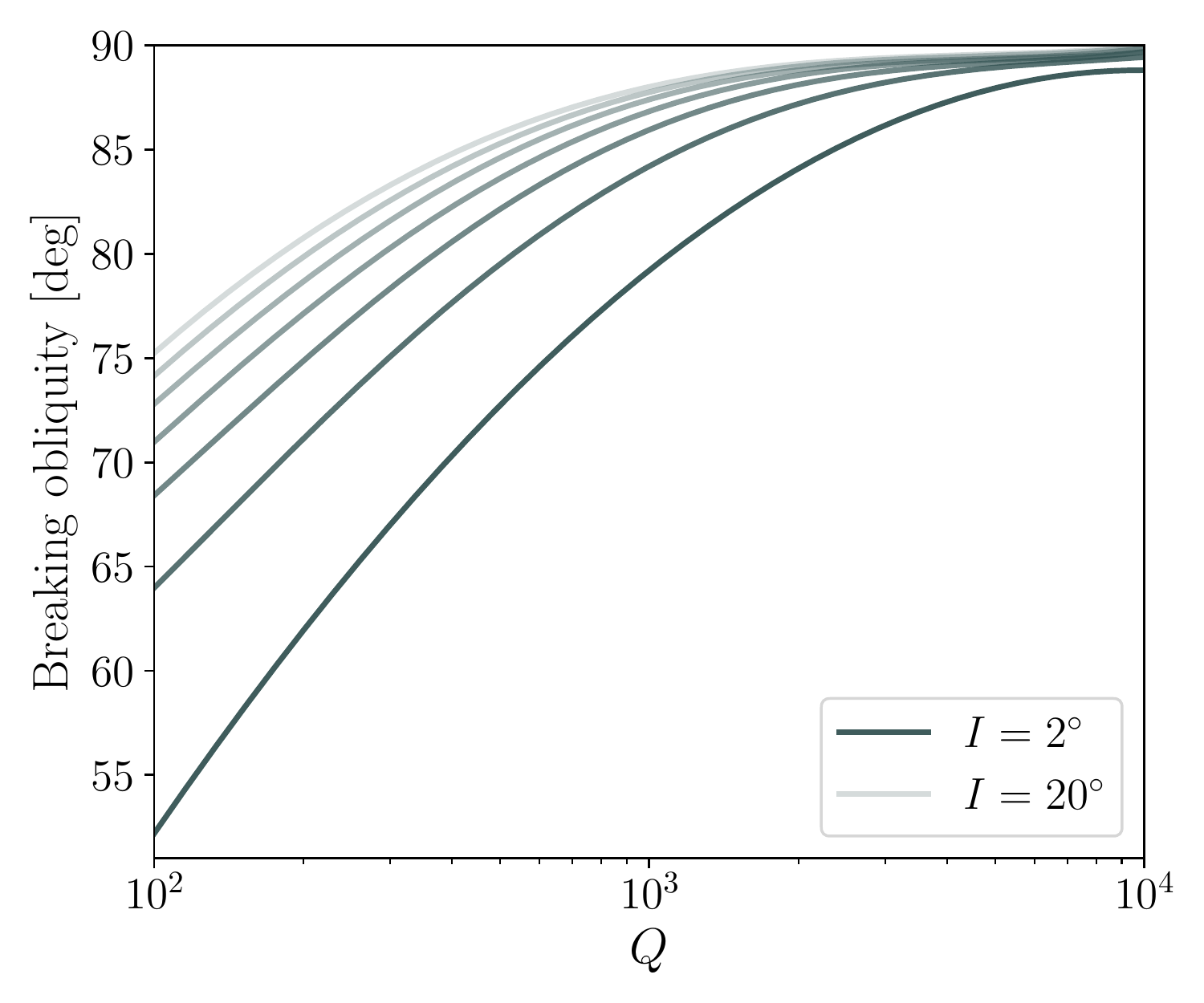}
\caption{Breaking obliquity of Cassini state 2 (maximum obliquity that is stable to tidal dissipation) as a function of $Q$ and $I$. Each curve corresponds to a different $I$, with the darkest curve being $I=2^{\circ}$ and the lightest curve being $I=20^{\circ}$. There is a change of $\Delta I = 3^{\circ}$ between each curve.} 
\label{fig: Breaking_obliquity}
\end{figure}

For the purposes of this calculation, we will assume the normal vector $\bm{\hat{n}}$ precesses uniformly about the total angular momentum vector $\bm{\hat{k}}$ with a constant inclination $I$ between them. We will initialize the system with $|g|/{\alpha} = 3 > (|g|/{\alpha})_{\mathrm{crit}}$ and let $|g|/{\alpha}$ exponentially decrease with a timescale equal to ten times the adiabatic limit given by \cite{2020arXiv200414380S}, i.e. firmly in the adiabatic regime. 
%This mimics the evolution of $|g|/{\alpha}$ due to orbital decay of the innermost planet.
The planet starts in Cassini state 2 with an obliquity close to $\epsilon\sim I$ (see Figure~\ref{fig: Four_cassini_states}). As $|g|/{\alpha}$ decreases, the obliquity rises up the Cassini state 2 branch until the dissipative torque becomes too strong, and the planet can no longer be maintained in the high obliquity state. 
%The spin vector breaks and quickly damps down to Cassini state 1. 
Figure~\ref{fig: Breaking_obliquity} shows the numerically-calculated breaking obliquity for a range of values of $Q$ and $I$. We observe that the limit increases with both $Q$ and $I$, indicating that planets with such properties can undergo more orbital decay before tidal breaking.

\begin{comment}
This is the analytical solution from Fabrycky et al. 

The breaking point can be identified by considering the limits of a phase shift that appears in Cassini state 2 in the presence of steady tidal dissipation. The spin axis, $\bm{\hat{s}}$, shifts slightly out of the plane defined by $\bm{\hat{n}}$ and $\bm{\hat{k}}$, and this shift provides a balance of the tidal torque (up until the breaking point). The phase shift, $\phi$, obeys the relation,
\begin{equation}
\sin(\delta+\phi)\csc\delta = \tan\epsilon\cot{I},
\label{eq: phase shift relation}
\end{equation}
%\begin{equation}
%\phi \approx (\xi\cos\epsilon)^{-1}(\tan\epsilon\cot I -1)
%\end{equation}
where
\begin{equation}
\begin{split}
\xi &\equiv \frac{2Q}{3}\frac{\omega}{n} \\
\cot\delta &\equiv \xi\cos\epsilon.
\end{split}
\end{equation}
The phase shift increases with the obliquity, but it cannot go beyond $\phi = \pi/2$, which is the point at which Cassini state 2 is destabilized. Substituting $\omega = \omega_{\mathrm{eq}}$ (equation \ref{eq: omega_eq}) and $\phi = \pi/2$ in equation \ref{eq: phase shift relation}, we obtain a relation for the obliquity, $\epsilon$, at the breaking point.
\begin{equation}
(\tan\epsilon\cot{I})(1 + \sec^2\epsilon) < \frac{4Q}{3}
\label{eq: breaking obliquity}
\end{equation}
\end{comment}

After Cassini state 2 is destabilized, the obliquity damps down from its excited state and settles into Cassini state 1. From equation \ref{eq: torques}, we see that this equilibration occurs on a fast timescale of roughly
\begin{equation}
\begin{split}
\tau_{\mathrm{equil}} &\approx t_f C \left(\frac{R_p}{a}\right)^2 \\
&= 135 \ \mathrm{yr}\Big(\frac{Q'}{10^3}\Big)\Big(\frac{a}{0.03 \mathrm{AU}}\Big)^{\frac{9}{2}}\Big(\frac{\rho_p}{\rho_{\oplus}}\Big)\Big(\frac{C}{0.35}\Big)\Big(\frac{M_{\star}}{M_{\odot}}\Big)^{-\frac{3}{2}}.
\label{eq: tau_equil}
\end{split}
\end{equation}
Once in Cassini state 1, the obliquity is small but non-zero (equation \ref{eq: Cassini state 1 and 2} and Figure \ref{fig: Four_cassini_states}). For instance, when $I = 5^{\circ}$ and $|g|/{\alpha} = 0.1$, the obliquity of Cassini state 1 is equal to $\epsilon_1 \approx -0.5^{\circ}$. The planet may experience further orbital decay while in Cassini state 1, but it will generally be slow and stable, since the Cassini state 1 obliquity decreases in magnitude as $|g|/{\alpha}$ decreases.

\section{Parameter Space for USP Production}
\label{sec: Parameter Space for USP Production}

%1) enter a Cassini state through tidal relaxation
%2) undergo tidal migration due to the non-zero obliquity
%3) eventually exit the migration due to the Cassini state becoming unstable

We have just outlined a mechanism by which a short-period planet can undergo full-scale orbital decay via obliquity-driven tidal migration. This process applies to a subset of planets that meet two criteria: (1) their tidally-relaxed spin states are forced to have non-zero obliquities; and (2) their initial semi-major axes are short enough to trigger tidal migration on a rapid timescale. These criteria must be concretely specified in terms of the planetary parameter space that is susceptible to USP production. To do this, we will first set up the system and identify its most relevant parameters (Section \ref{sec: System set-up and parameter space definition}). Next, we will define the secular orbital frequencies (Section \ref{sec: Calculation of g/alpha using secular frequencies}) and use these to delineate the parameter region that is susceptible to USP production (Section \ref{sec: Identifying the susceptible parameter space for USP production}). 

\subsection{System set-up and parameter space definition}
\label{sec: System set-up and parameter space definition}

We consider the innermost planet in a close-in, multi-planet system that is not perfectly coplanar. Mutual inclination constraints will be discussed in Section \ref{sec: Limitation: angular momentum budget}, but they do not matter in detail for now. In addition, we will adopt a three-planet system. This again does not strongly affect the overall picture. Although working with two-planet systems simplifies the dynamics, it is less generalizable, and three-planet systems are more representative of the observed multi-planet systems with USPs. 

In the process of mapping out the parameter space, there are many system properties to consider, including $M_p$, $R_p$, $a$, period ratios $P_{j+1}/P_j$ between neighboring planets, etc. We will reduce the exploration down to three essential parameters: $M_p$, $a_{1,i}$ (the innermost planet's initial semi-major axis), and $P_{j+1}/P_j$. We assume that all planets in the system have uniform masses and orbital spacing, a simplifying assumption that we will later relax (Section \ref{sec: Unequal planet masses}) but which is justified on the basis of the observed intra-system uniformity of \textit{Kepler} multis \citep{2018AJ....155...48W, 2017ApJ...849L..33M}. In addition, we assume that the inner planet has an Earth-like composition (approximately $30\%$ Fe, $70\%$ MgSiO$_3$), and we use this assumption to calculate $R_{p1}$ for a given $M_{p1}$  (where the subscript `1' refers to the innermost planet) using the mass-radius tables from \cite{2016ApJ...819..127Z}. (In Section \ref{sec: Early system evolution}, we will discuss the implications of possible early mass loss from the inner planet, which would be expected if the planet formed with a primordial envelope of hydrogen/helium.)

Apart from the three essential parameters ($M_p$, $a_{1,i}$, and $P_{j+1}/P_j$), there are several additional parameters that we will hold fixed. We will take the inner planet's Love number and moment of inertia factor (which enter into $\alpha$ in equation \ref{eq: alpha}) to represent fiducial values for terrestrial planets, as determined observationally for Solar System bodies \citep{1999ssd..book.....M, 2016CeMDA.126..145L} and theoretically for extrasolar super-Earths \citep{2011A&A...528A..18K, 2018A&A...615A..39K}. We will use $Q=10^3$, $k_2 = 0.4$, and $C = 0.35$, noting that changes in these parameters will only affect our results on a detailed level. We will also assume that the planet's spin rate is at equilibrium, $\omega=\omega_{\mathrm{eq}}$, or approximately synchronous with the mean-motion, $\omega = n$, when eccentricities and obliquities are small.\footnote{This assumption is appropriate because the synchronization timescale, $\tau_{\mathrm{sync}} \approx \omega/{\dot{\omega}}$, which is the same as $\tau_{\mathrm{equil}}$ in equation \ref{eq: tau_equil}, is only $\sim 10^2-10^3$ yr for planets with $a\lesssim0.05$~AU.}
As for mutual inclinations, we will take the innermost planet to be misaligned by $10^{\circ}$ with respect to the outer planets (approximately consistent with \citealt{2018ApJ...864L..38D}), with a $1^{\circ}$ mutual inclination between the outer planets. Finally, we will consider two different stellar masses, $M_{\star} = 0.6 \ M_{\odot}$ (K-dwarf star) and $M_{\star} = 1.0 \ M_{\odot}$ (G-dwarf star). 

Before proceeding, we note that evolution by way of obliquity tides must conserve angular momentum, in spite of orbital decay. Accordingly, the inclination of the inner planet must change as its orbit shrinks \citep{2007ApJ...665..754F}. Such inclination evolution does not affect this section substantially, since Cassini state 2 varies little with inclination at the high obliquity end (Figure \ref{fig: Four_cassini_states}). However, the inclination evolution is important to incorporate into our theory because angular momentum constraints can limit the total extent of the migration. Thus, we redress this omission in Section \ref{sec: Limitation: angular momentum budget} and Appendix \ref{sec: Secular model with tides}, where we develop a secular model that self-consistently accounts for the tidal semi-major axis and inclination evolution.

\subsection{Calculation of $|g|/{\alpha}$ using secular frequencies}
\label{sec: Calculation of g/alpha using secular frequencies}
The primary factor determining whether the innermost planet will become a USP is its Cassini state, and this depends most strongly on $|g|/{\alpha}$. Accordingly, our goal is to calculate this frequency ratio for the innermost planet across the parameter space we have just identified. While the spin-axis precession constant $\alpha$ has a straightforward analytic expression (equation \ref{eq: alpha}), $g$ is more complex. The planet's orbit nodal precession arises from gravitational interactions with its (potentially oblate) host star and neighboring planets. Assuming the orbits are non-resonant, the set of orbital eigenfrequencies, $\{g_i\}$, may be approximated using Laplace-Lagrange secular theory.\footnote{For closely-spaced planets with several degree mutual inclinations and eccentricities summing to $\sim0.1$, the nodal precession frequency can differ from Laplace-Lagrange by $\sim10\%$ or even greater \citep{2020AJ....159..217B}. While this is significant, the range in possible $\alpha$ values (due to unknown $k_2$ and $C$) is at a similar level. Accordingly, for simplicity, we adopt Laplace-Lagrange secular frequencies throughout this work.} To second order in the eccentricities and inclinations, this solution depends only on the masses and semi-major axes, and the eccentricity and inclination solutions are decoupled. As discussed in Section \ref{sec: Entry into Cassini states}, the relevant nodal frequency $g$ for the Cassini state will be one of the multiple eigenfrequencies identified in this solution.

We begin by constructing a planetary disturbing function for $N$ planets orbiting an oblate host star \citep{1999ssd..book.....M}. The disturbing function is the non-Keplerian perturbing gravitational potential experienced by the planets due to their mutual interactions. Keeping only terms associated with the inclinations to second order, the disturbing function takes the form

\begin{equation}
\left< \mathcal{R}_j^{\mathrm{(sec)}} \right> = n_j a_j^2\Bigg[\frac{1}{2}B_{jj}I_j^2 + \sum_{\substack{k=1 \\ k \ne j}}^{N} B_{jk} I_j I_k \cos(\Omega_j - \Omega_k) \Bigg],   
\label{eq: disturbing function}
\end{equation}
where the subscript $j$ is the planet number, $n$ is the mean motion, $I$ is the inclination, and $\Omega$ is the longitude of the ascending node. The quantities $B_{jj}$ and $B_{jk}$ represent the interaction coefficients within the matrix $\bm{B}$ and are given by
\begin{align}
B_{jj} &= -n_j\Bigg[\frac{3}{2}J_{2\star}\left(\frac{R_{\star}}{a_j}\right)^2 - \frac{27}{8}J_{2\star}^2\left(\frac{R_{\star}}{a_j}\right)^4 \label{eq: B_jj} \\
&+ \frac{1}{4}\sum_{\substack{k=1 \\ k \ne j}}^{N}\frac{M_{pk}}{M_{\star} + M_{pj}}\alpha_{jk}\bar{\alpha}_{jk}b_{3/2}^{(1)}(\alpha_{jk})\Bigg] \nonumber \\
%\end{align}
%\begin{equation}
B_{jk} &= \frac{1}{4}\frac{M_{pk}}{M_{\star} + M_{pj}}n_j\alpha_{jk}\bar{\alpha}_{jk}b_{3/2}^{(1)}(\alpha_{jk}) \ \ \ (j \ne k).
\end{align}
Here, $M_{pj}$ is the mass of the $j$th planet. When $a_j < a_k$, $\alpha_{jk} = \bar{\alpha}_{jk} = a_j/a_k$, and when $a_j > a_k$, $\alpha_{jk} = a_k/a_j$ and $\bar{\alpha}_{jk} = 1$. The quantity $b_{3/2}^{(1)}$ is a Laplace coefficient, defined by
\begin{equation}
b_{3/2}^{(1)}(\alpha) = \frac{1}{\pi}\int_0^{2\pi}\frac{\cos\psi \ \mathrm{d}\psi}{(1-2\alpha\cos\psi + \alpha^2)^{3/2}}.
\end{equation}

In equation \ref{eq: B_jj}, $J_{2\star}$ is the star's second gravitational (quadrupole) moment. $J_{2\star}$ can be expressed in terms of $R_{\star}$, the stellar spin rate, $\omega_{\star} = 2\pi/P_{\star}$, and the tidal Love number $k_{2\star}$ using the approximate relationship \citep{1939MNRAS..99..451S, 1976Icar...28..441W, 2016ApJ...830....5S}
\begin{align}
\label{eq: J2}
J_{2\star}&\approx \frac{1}{3}k_{2\star}\frac{\omega_\star^2}{G M_\star/R_\star^3}  \\
&\sim 10^{-3}\bigg(\frac{k_{2\star}}{0.2}\bigg)\bigg(\frac{P_\star}{\textrm{day}}\bigg)^{-2}\bigg(\frac{R_\star}{R_\odot}\bigg)^3\bigg(\frac{M_\star}{M_\odot}\bigg)^{-1}, \nonumber
\end{align}
where $\sqrt{G M_\star/R_\star^3}$ is the break-up rotational velocity. Here we have used fiducial values appropriate to young, rapidly-rotating stars \citep{2013ApJ...778..169B, 2016ApJ...830....5S}. As we will show, the inclusion of $J_{2\star}$ is not required for the mechanism but does make it more efficient. Going forward, unless otherwise noted, we will use $J_{2\star} = 10^{-4}$ to represent a typical star within the first several 100 Myrs of evolution. 

Given the form of the disturbing function in equation \ref{eq: disturbing function}, it is convenient and customary to use a transformation to the inclination ``vectors'', defined by
\begin{equation}
\begin{split}
p_j &= I_j\sin\Omega_j \\
q_j &= I_j\cos\Omega_j. 
\end{split}
\end{equation}
The solutions to the equations of motion are then
\begin{equation}
\begin{split}
p_j(t) &= \sum_{i=1}^{N}I_{ji}\sin(g_i t + \gamma_i) \\
q_j(t) &= \sum_{i=1}^{N}I_{ji}\cos(g_i t + \gamma_i),
\label{eq: p and q}
\end{split}
\end{equation}
where the $\{g_i\}$ are the $N$ eigenvalues of the matrix $\bm{B}$, and $\{I_{ji}\}$ are the corresponding eigenvectors.\footnote{We note that \cite{1999ssd..book.....M} use $g_i$ and $f_i$ to denote the eigenfrequencies corresponding to the eccentricity and inclination solutions, respectively. Here we use $g_i$ to denote the inclination eigenfrequencies in order to maintain consistency with the notation of standard literature on Cassini states.} Since the eigenvectors of $\bm{B}$ are only defined up to a scaling factor, one may use the initial conditions to determine the magnitudes of the eigenvectors and the phases $\gamma_i$. Finally, the time evolutions of the inclination and node are given by 
\begin{equation}
\begin{split}
I_j(t) &= \left[[p_j(t)]^2 + [q_j(t)]^2\right]^{1/2} \\
\Omega_j(t) &= \tan^{-1}\left[\frac{p_j(t)}{q_j(t)}\right].
\label{eq: inclination solution}
\end{split}
\end{equation}

Equations \ref{eq: p and q} and \ref{eq: inclination solution} indicate that the inclination/node solution is composed of a superposition of modes from the $N$ secular eigenfrequencies, $\{g_i\}$. Any of these modes may be the $g$ frequency that is dominant for a planet's Cassini state, such as in the case of Saturn, where the relevant $g$ is that which is dominated by Neptune's nodal precession \citep{2004AJ....128.2501W, 2004AJ....128.2510H}. Determining the important $\{g_i\}$ mode is challenging, but for our regime of interest it will generally be the one that is closest to the spin-axis precession constant $\alpha$. In the sections that follow, we will take the fastest frequency $|g|_{\mathrm{max}}$ as the dominant mode. We will show in Section \ref{sec: initial system evolution} that this is a good approximation. A robust determination of which of the $\{g_i\}$ modes dominates is the biggest area for follow-up of this work. This would include an investigation of when and how chaos arises from overlapping modes. We will return to this idea in the Discussion (Section \ref{sec: Discussion}). 

\subsection{Identifying the susceptible parameter space for USP production}

\begin{figure}
\epsscale{1.}
\plotone{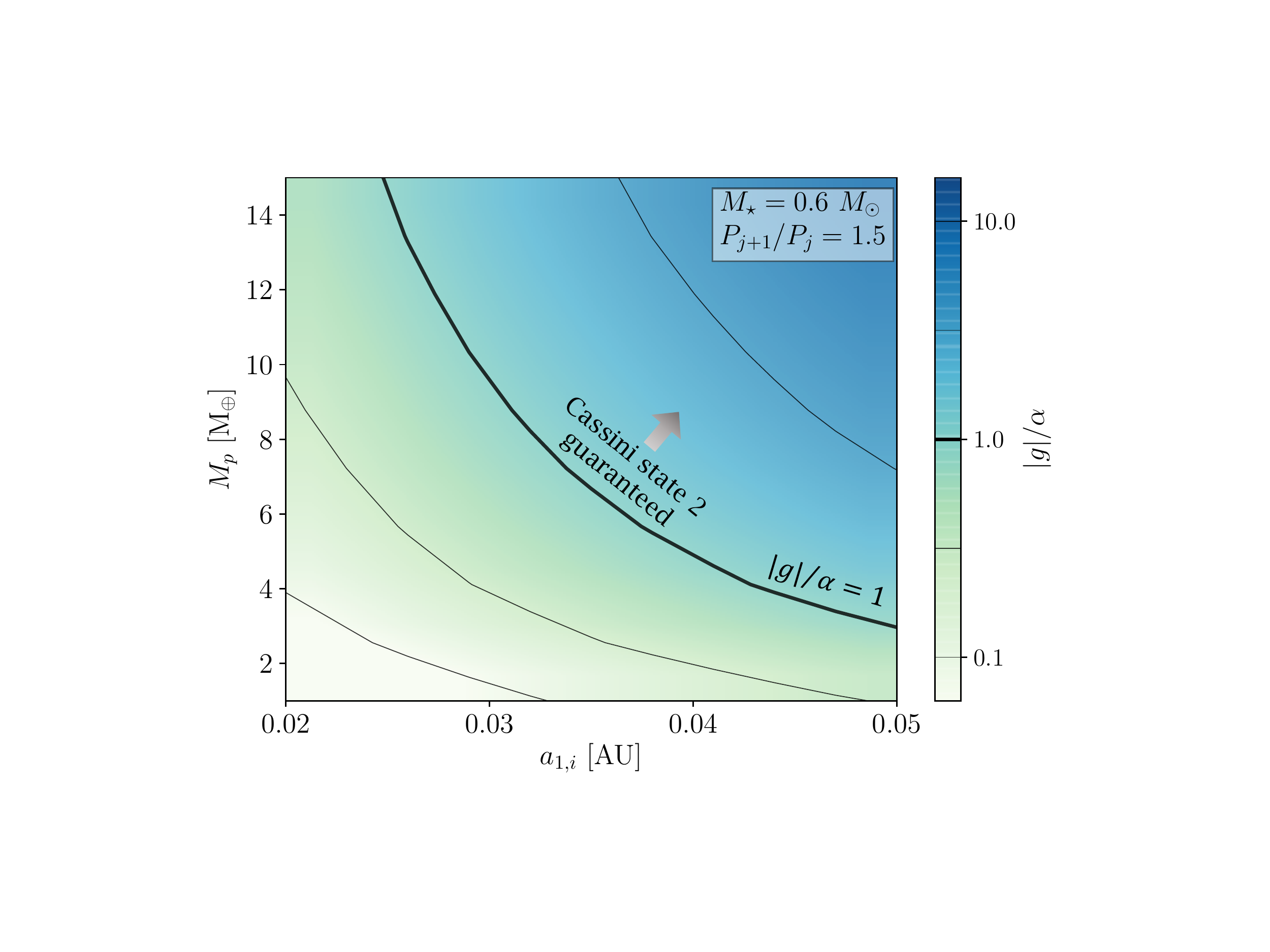}
\caption{Variation of $|g|/{\alpha}$ as a function of $M_p$ and $a_{1,i}$ for a fixed stellar mass, $M_{\star} = 0.6 \ M_{\odot}$ and period ratio, $P_{j+1}/P_j = 1.5$. The thick contour line corresponds to $|g|/{\alpha} = 1 \gtrsim (|g|/{\alpha})_{\mathrm{crit}}$. When $|g|/{\alpha} > 1$ and the spin direction is prograde, tidally-induced capture into Cassini state 2 is guaranteed. } 
\label{fig: g_alpha_heatmap}
\end{figure}

With the $M_p$ -- $a_{1,i}$ -- $P_{j+1}/P_j$ parameter space and the calculation of $|g|/{\alpha}$ now specified, we can identify the region of this space that is susceptible to producing a USP via obliquity-driven tidal migration. To begin, we simply plot in Figure \ref{fig: g_alpha_heatmap} $|g|/{\alpha}$ as a function of $M_p$ and $a_{1,i}$ for a fixed $P_{j+1}/P_j$. There are several aspects to note. First, for a fixed $M_p$, the ratio $|g|/{\alpha}$ increases with $a_{1,i}$, while both $|g|$ and $\alpha$ independently \textit{decrease} with increasing $a_{1,i}$. This highlights the fact that $|g|/{\alpha}$ will shrink upon the inner planet's orbital decay, as depicted in Figure \ref{fig: Four_cassini_states}. (During the decay, however, the period ratio is also increasing, which leads to a more rapid decrease of $|g|/{\alpha}$ than for a fixed $P_{j+1}/P_j$.) 

\label{sec: Identifying the susceptible parameter space for USP production}

\begin{figure}
\epsscale{1.2}
\plotone{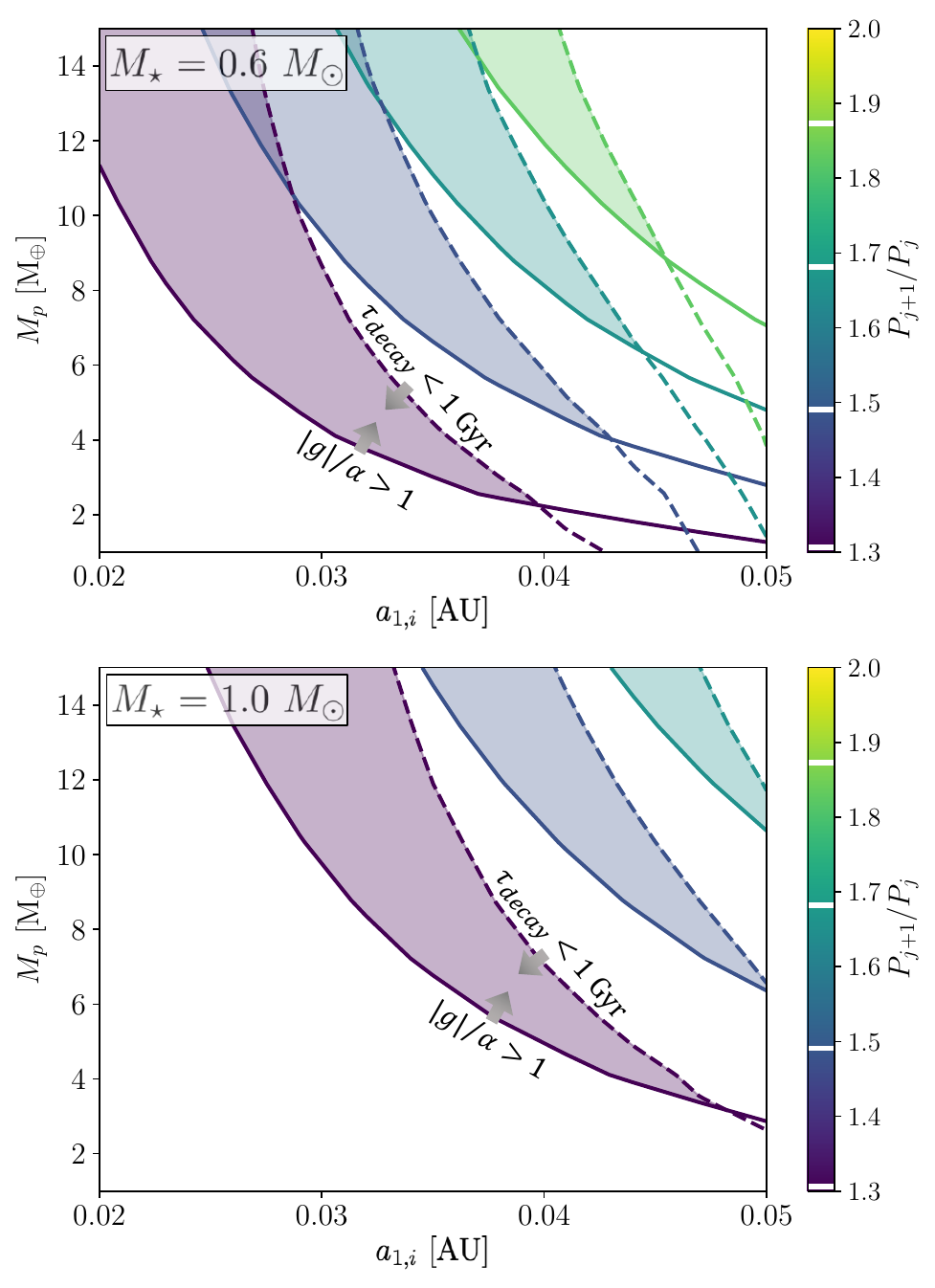}
\caption{Allowable parameter space for USP formation. As a function of $M_p$ and $a_{1,i}$, the solid lines show the $|g|/{\alpha} = 1$ contours (c.f. the thick contour line in Figure \ref{fig: g_alpha_heatmap}) for a range of period ratios, $P_{j+1}/P_j$, indicated with the colorbar and the white horizontal lines. The top and bottom panels are identical except for the stellar mass, $M_{\star} = 0.6 \ M_{\odot}$ (top) and $M_{\star} = 1.0 \ M_{\odot}$ (bottom). As in Figure \ref{fig: g_alpha_heatmap}, $|g|/{\alpha}$ increases towards the upper right. The dashed lines are the 1 Gyr contours of $\tau_{\mathrm{decay}}$, which decreases towards the left. This creates the shaded regions with $|g|/{\alpha} > 1$ and $\tau_{\mathrm{decay}} < 1$ Gyr where USP formation can occur. } 
\label{fig: g_alpha=1_contours_annotated}
\end{figure}

\begin{figure}
\epsscale{1.2}
\plotone{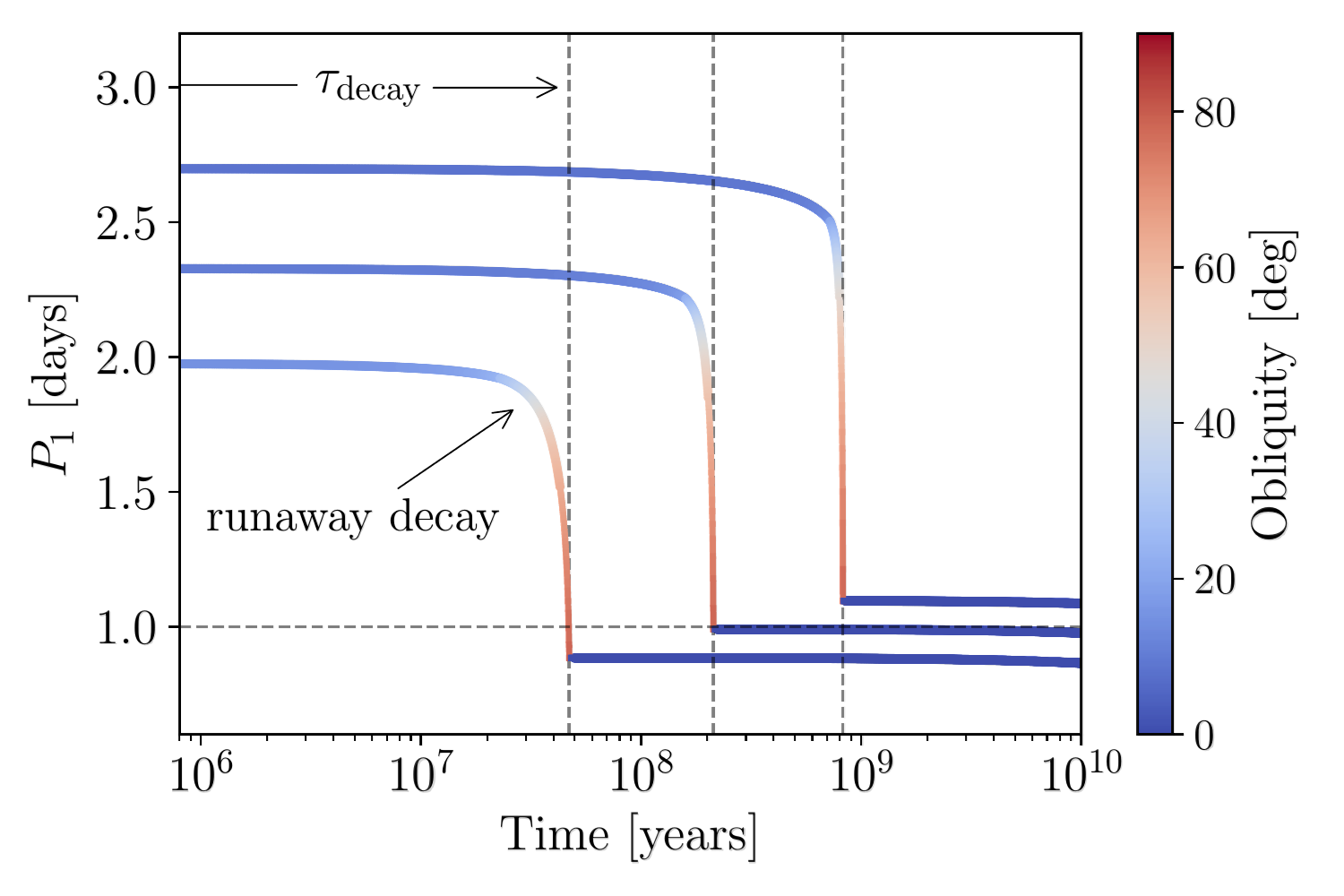}
%Mp_list = [7.22, 8.78, 10.33]
%a1_list = [0.026, 0.029, 0.032]
%PR_list = [1.3, 1.364, 1.427]
\caption{Three example time evolution curves of the innermost planet's orbital period and obliquity. These correspond to three starting grid points in the $M_p$ -- $a_{1,i}$ -- $P_{j+1}/P_j$ parameter space, so their parameters are arbitrary. From bottom curve to top, the parameters are $M_p = [7.22, 8.78, 10.33] \ M_{\oplus}$, $a_1 = [0.026, 0.029, 0.032] \ \mathrm{AU}$, and  $P_{j+1}/P_j = [1.3, 1.364, 1.427]$, with $M_{\star} = 0.6 \ M_{\odot}$ for all three. The vertical dashed lines represent $\tau_{\mathrm{decay}}$ for each curve, and the horizontal dashed line at $P_1=1$ day simply indicates the USP cutoff. The obliquity begins in Cassini state 2 with a low value and increases as a result of the inward migration. The planet then reaches an epoch of rapid, runaway orbital decay. This phase is stalled when the obliquity reaches the tidal breaking limit of Cassini state 2 and damps back down to Cassini state 1. The time, $\tau_{\mathrm{decay}}$, to reach the end of the rapid decay depends primarily on the starting semi-major axis, but also on $Q$, $P_{j+1}/P_j$, etc.} 
\label{fig: Secular_orbital_decay_examples}
\end{figure}

A second observation is that the $|g|/{\alpha} = 1$ contour crosses through the middle of this parameter space. For $|g|/{\alpha} > (|g|/{\alpha})_{\mathrm{crit}} \approx 1$, the inner planet is guaranteed to occupy Cassini state 2 (see Figure \ref{fig: Four_cassini_states}). This is not to say that the planet can't initially be captured into Cassini state 2 if $|g|/{\alpha} < (|g|/{\alpha})_{\mathrm{crit}}$. For $|g|/{\alpha} < (|g|/{\alpha})_{\mathrm{crit}}$ and when the obliquity is greater than the dashed separatrix curves in Figure \ref{fig: Four_cassini_states}, the obliquity tidally relaxes into state 2. In addition, the obliquity can also be resonantly excited into state 2 \citep[e.g.][]{2019NatAs...3..424M}. However, in this work, we focus on the $|g|/{\alpha} > (|g|/{\alpha})_{\mathrm{crit}} \approx 1$ regime where capture into Cassini state 2 is inevitable.

Given that the $|g|/{\alpha}=1$ contour delineates the region of guaranteed participation in Cassini state 2, we can isolate this contour and plot its variation with respect to the third parameter, $P_{j+1}/P_j$, which was held fixed in Figure \ref{fig: g_alpha_heatmap}. The result is shown in Figure \ref{fig: g_alpha=1_contours_annotated}, where the solid lines indicate the $|g|/{\alpha}=1$ contours for a range of $P_{j+1}/P_j$ given by the colorbar. For increasing $P_{j+1}/P_j$, $|g|$ decreases, so the $|g|/{\alpha}=1$ contour moves to larger $a_{1,i}$.

In addition to the $|g|/{\alpha}\gtrsim 1$ constraint, USP production also requires that $a_{1,i}$ is small enough such that $\tau_{\mathrm{decay}}$ is sufficiently fast (i.e. $\lesssim 1$ Gyr). To calculate $\tau_{\mathrm{decay}}$, we can integrate $\dot{a}$ from equation \ref{eq: tau_a} while asserting that the obliquity $\epsilon$ is in a Cassini state solved numerically using equation \ref{eq: Cassini state relation}. The obliquity begins in Cassini state 2, $\epsilon = \epsilon_2$. As discussed in Section \ref{sec: Obliquity-driven tidal migration}, the orbital migration leads $\epsilon_2$ to increase, and the decay rate eventually reaches runaway. Accordingly, $\tau_{\mathrm{decay}}$ is the time it takes until the tidal breaking of Cassini state 2 (recall Section \ref{sec: Tidal breaking of Cassini states}). The obliquity subsequently settles into Cassini state 1, $\epsilon = \epsilon_1$, over a short timescale ($\sim 100$ yr, equation \ref{eq: tau_equil}) that we approximate as instantaneous. Finally, in this low obliquity state, the rapid orbital decay stalls.  

For each point in the $M_p$ -- $a_{1,i}$ -- $P_{j+1}/P_j$ grid, we integrate $\dot{a}$ from equation \ref{eq: tau_a} over 10 Gyr and calculate $\tau_{\mathrm{decay}}$ as just described. Examples for three initial grid points are shown in Figure \ref{fig: Secular_orbital_decay_examples}, where we plot the time evolution of $P_1$ and $\epsilon$. All three examples undergo a period of runaway decay, although they reach different final orbital periods at different $\tau_{\mathrm{decay}}$ times. For similar initial period ratios, these outcomes depend most strongly on the initial semi-major axes, $a_{1,i}$. 

As a result of performing these orbital decay evolutions for all grid points, the dashed lines in Figure \ref{fig: g_alpha=1_contours_annotated} indicate the $\tau_{\mathrm{decay}} = 1$ Gyr contours for a range of $P_{j+1}/P_j$ values. For a given contour, the area to the left corresponds to shorter $\tau_{\mathrm{decay}}$. Note that, for larger $P_{j+1}/P_j$ at fixed $a_{1,i}$ and $M_p$, the orbital precession frequency $|g|$ is smaller, such that the initial $\epsilon_2$ is larger and $\tau_{\mathrm{decay}}$ is smaller. This accounts for the observation that the $\tau_{\mathrm{decay}} = 1$ Gyr contours shift to the right for larger $P_{j+1}/P_j$. Similarly, comparing the top and bottom panels of Figure \ref{fig: g_alpha=1_contours_annotated}, we see that all contours are further to the right (larger $a_{1,i}$) for the $M_{\star} = 1.0 \ M_{\odot}$ case compared to the $M_{\star} = 0.6 \ M_{\odot}$ case. This is related to the fact that, with all other parameters held fixed, a larger $M_{\star}$ yields a smaller $|g|/{\alpha}$ and $\tau_{\mathrm{decay}}$. 

Taken together, the regions with $|g|/{\alpha} > 1$ and $\tau_{\mathrm{decay}} < 1$ Gyr delineate the parts of parameter space (shaded in Figure \ref{fig: g_alpha=1_contours_annotated}) where the innermost planets are most susceptible to becoming a USP. (We note that these regions have been plotted for the illustrative set of period ratios and thus do not indicate the full parameter space.) It is intriguing to note that there is a larger susceptible parameter space (meaning more frequent USP production) for the case with $M_{\star} = 0.6 \ M_{\odot}$ compared to that with $M_{\star} = 1.0 \ M_{\odot}$. This trend is in the same direction as the observational occurrence rates derived by \cite{2014ApJ...787...47S}, who found that USPs are more common around smaller mass stars. While this comparison is suggestive, we would also need to know the planet occurrence rate as a function of $M_p$, $a$, and $M_{\star}$ to robustly determine that the theory has made a correct prediction.

\subsubsection{Unequal planet masses}
\label{sec: Unequal planet masses}
In our analysis thus far, we have adopted the simplifying assumption of equal mass planets in a compact, equally-spaced system. This configuration is not always a good approximation, however, and it is useful to consider systems hosting more massive planets on wider exterior orbits. Still adopting a three-planet system, we keep the innermost planet's mass, $M_{p1}$, fixed and examine a range of masses and separations for the two exterior planets. We parameterize this using $M_{p,\mathrm{ext}}/M_{p1}$ (where $M_{p,\mathrm{ext}}$ is the mass of each exterior planet) and $P_{j+1}/P_j$. 

\begin{figure}
\epsscale{1.2}
\plotone{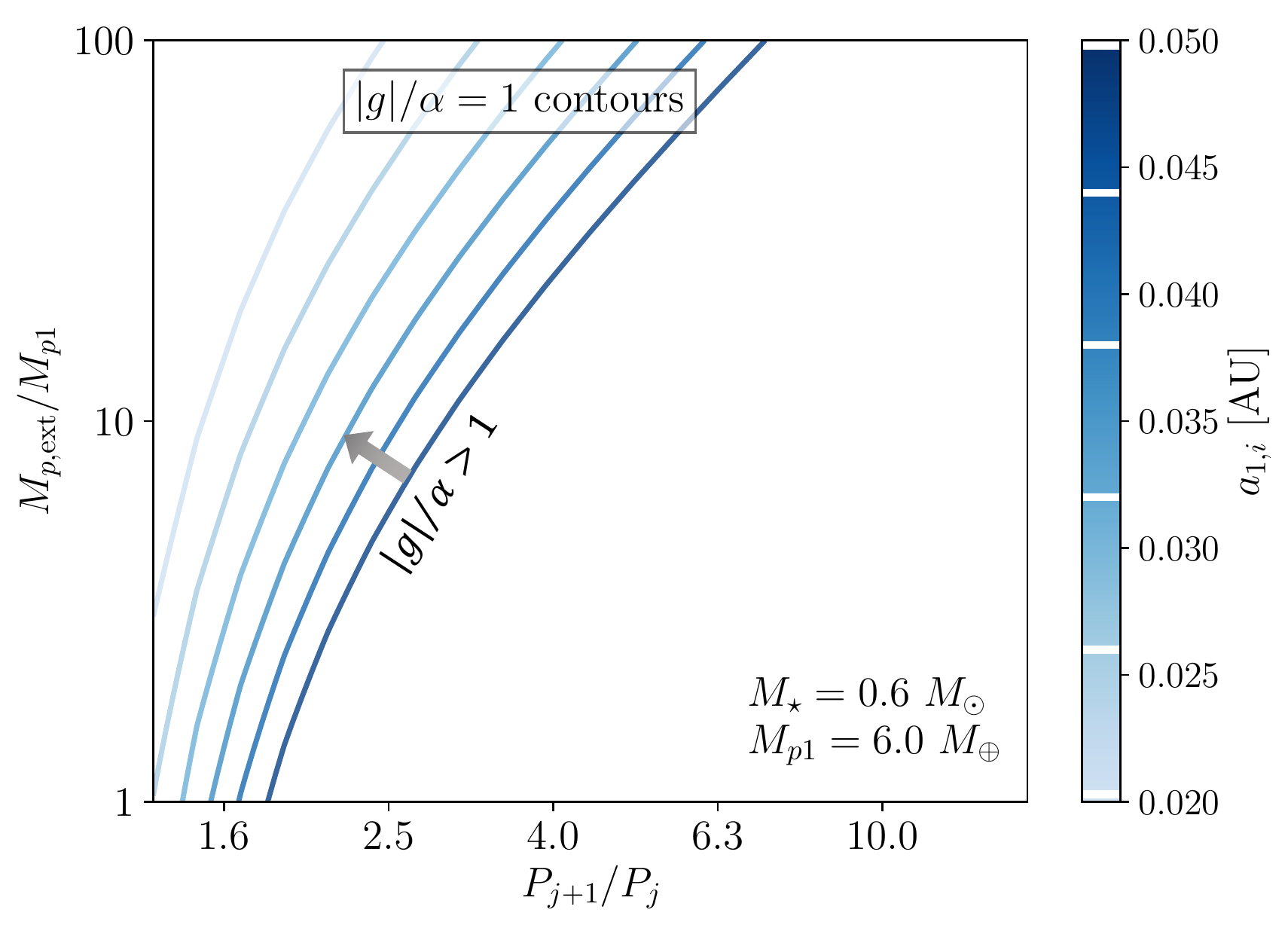}
\caption{Contours of $|g|/{\alpha} = 1$ for the innermost planet after relaxing the simplifying assumption of equal mass planets. The contours are plotted as a function of the mass ratio between the outer and inner planets, $M_{p,\mathrm{ext}}/M_{p1}$, and the period ratio, $P_{j+1}/P_j$. Each curve corresponds to a different value of $a_{1,i}$ (represented by the colorbar) but with a fixed mass for the innermost planet, $M_{p1} = 6 \ M_{\oplus}$. For a given contour, the region of $|g|/{\alpha} > 1$ is towards the upper left. Larger exterior planets at wider separations can have a similar dynamical effect as equal-mass perturbers at close separations, in the sense that they produce similar orbital precession rates, $|g|$. } 
\label{fig: g_alpha=1_contours_Mp_Mp1_vs_PR_variation}
\end{figure}

Figure \ref{fig: g_alpha=1_contours_Mp_Mp1_vs_PR_variation} shows contours of $|g|/{\alpha} = 1$ in the $M_{p,\mathrm{ext}}/M_{p1}$ -- $P_{j+1}/P_j$ space for $M_{p1} = 6 \ M_{\oplus}$ and for different values of $a_{1,i}$. (Note that the $|g|/{\alpha}$ ratio depends much more strongly on $a_{1,i}$ than $M_{p1}$, so varying $M_{p1}$ does not change this picture much.) The contours illustrate that more massive exterior planets with wider separations can have the same effect on the orbital precession frequency as smaller and closer perturbers. Even for $M_{p,\mathrm{ext}}/M_{p1} \sim 100$, however, $|g|/{\alpha}>1$ requires that $P_{j+1}/P_j < 10$.

\subsubsection{Initial obliquity evolution}
\label{sec: initial system evolution}

\begin{figure}
\epsscale{1.2}
\plotone{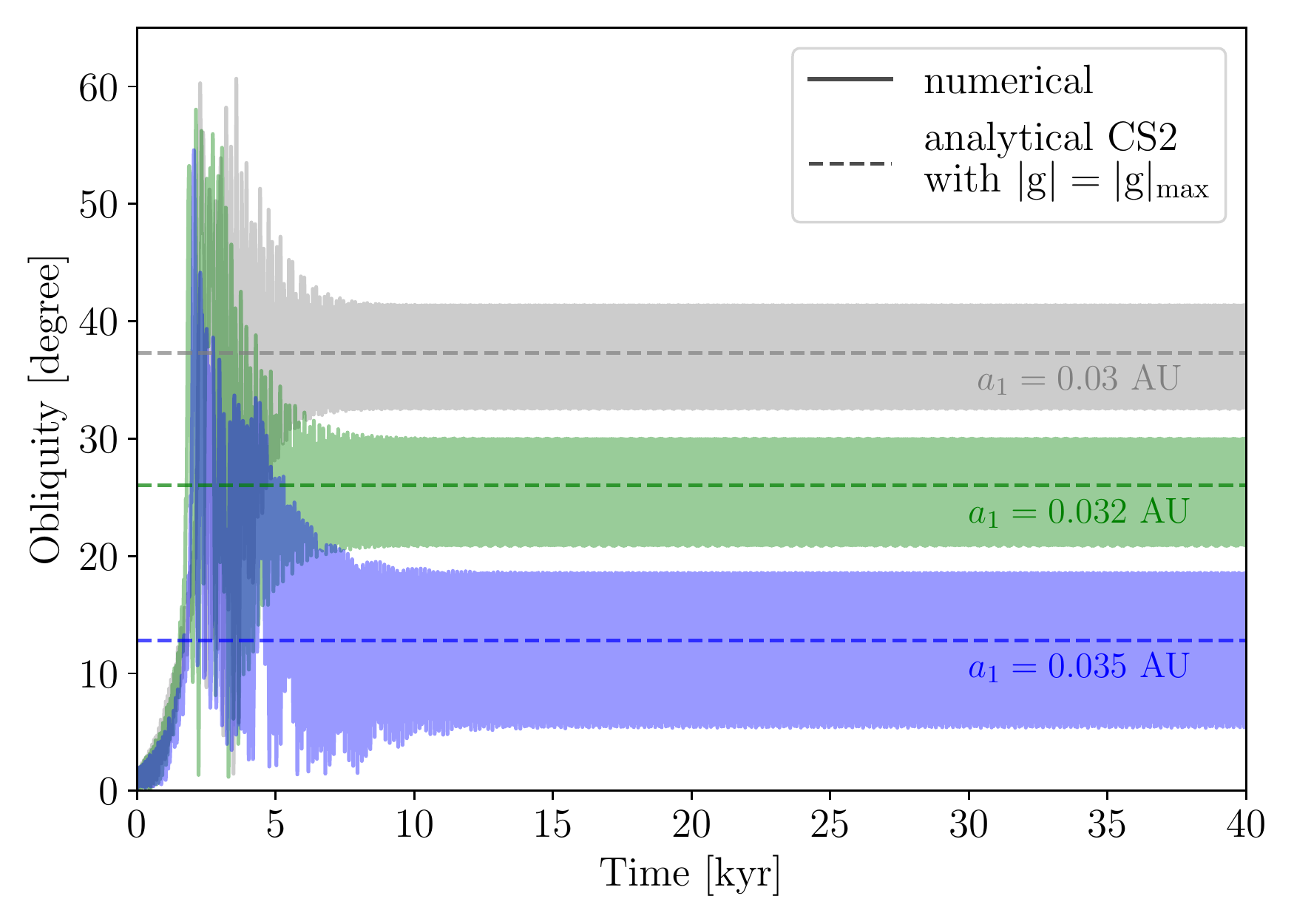}
\caption{Examples of the initial obliquity evolution showing capture into Cassini state 2. We show three different examples using the same masses, $M_p = 6 \ M_{\oplus}$, and period ratios, $P_{j+1}/P_j = 1.4$, but different semi-major axes for the innermost planet: $a_1 = 0.03$ AU (gray), $a_1 = 0.032$ AU (green), and $a_1 = 0.035$ AU (blue). The solid lines result from the numerical evolution of equation~\ref{eq: torque}, and the dashed line represents the analytical Cassini state 2 (CS2) obliquity calculated from equation~\ref{eq: Cassini state relation} using the fastest secular eigenfrequency, $|g|_{\mathrm{max}}$, and the corresponding eigenmode amplitude for $I$. The obliquities start at $\epsilon=0^{\circ}$. After a short period of chaotic evolution, they tidally relax into libration about Cassini state 2.} 
\label{fig: Initial_spin_vector_evolution_examples}
\end{figure}

We have assumed throughout Section \ref{sec: Identifying the susceptible parameter space for USP production} that the fastest frequency $|g|_{\mathrm{max}}$ is the dominant of the $\{g_i\}$ modes, but we have not yet justified this claim. Here we conduct a numerical study of the innermost planet's initial obliquity evolution in order to examine which mode is dominant early on and which Cassini state the planet initially settles into. We take the secular orbital solution (e.g. equation \ref{eq: inclination solution}) to represent the inclination and node evolution and evolve the spin vector using the secular equations of motion (equations \ref{eq: torque} and \ref{eq: torques}). We perform these evolutions for the same $M_p$ -- $a_{1,i}$ -- $P_{j+1}/P_j$ grid of initial conditions presented earlier. We consider a $0^{\circ}$ initial planet obliquity and a $0.5$ day primordial planetary rotation period. We also assume $Q_1 = 300$.  

Figure \ref{fig: Initial_spin_vector_evolution_examples} shows three examples of the innermost planet's initial spin vector evolution. Here we use $M_p = 6 \ M_{\oplus}$ and $P_{j+1}/P_j = 1.4$, but we show the results for different semi-major axes for the innermost planet. In the three examples, all of which have $|g|_{\mathrm{max}}/\alpha > (|g|/{\alpha})_{\mathrm{crit}}$, the obliquity starts from $0^{\circ}$, undergoes a transient period of chaotic excitation, and settles into libration around Cassini state 2 with $|g| = |g|_{\mathrm{max}}$ (the fastest secular eigenfrequency).

After performing these integrations for the full $M_p$ -- $a_{1,i}$ -- $P_{j+1}/P_j$ grid, we find good agreement with the analytic parameter boundaries identified earlier in Section \ref{sec: Identifying the susceptible parameter space for USP production}. That is, whenever $|g|_{\mathrm{max}}/\alpha \gtrsim 1$ (as represented with the solid contours in Figure \ref{fig: g_alpha=1_contours_annotated}), the obliquity is \textit{always} captured into Cassini state 2. About half the time, the Cassini state 2 is with the fastest frequency ($|g| = |g|_{\mathrm{max}}$), and the other times, it is with the second fastest frequency. Cassini state 2 with the second fastest frequency has a higher obliquity than that with $|g|_{\mathrm{max}}$. Accordingly, even if the spin vector temporarily settles into libration around Cassini state 2 with the second fastest frequency and later breaks out of it, it may still encounter Cassini state 2 with $|g|_{\mathrm{max}}$ as the obliquity damps back down. Finally, we also find that whenever $|g|_{\mathrm{max}}/\alpha \lesssim 1$, the obliquity is always captured into Cassini state 1 with $|g| = |g|_{\mathrm{max}}$. 

The results of these integrations thus confirm that the analytical simplification of using $|g|_{\mathrm{max}}$ as the dominant mode is appropriate. Cassini state 2 with $|g| = |g|_{\mathrm{max}}$ is a frequent occurrence whenever $|g|_{\mathrm{max}}/\alpha > 1$. However, these secular modes may not be readily distinguishable if the system parameters lead to resonance overlap. In this case, the planetary spin-vector would be susceptible to chaotic evolution (see Section \ref{sec: Obliquity chaos}). Further work is therefore required to better understand which initial state is most likely for arbitrary starting parameters. 

\section{Observed USP Planets in Multi-transiting Systems}
\label{sec: Observed USP Planets in Multi-transiting Systems}

\begin{figure*}
\epsscale{1.0}
\plotone{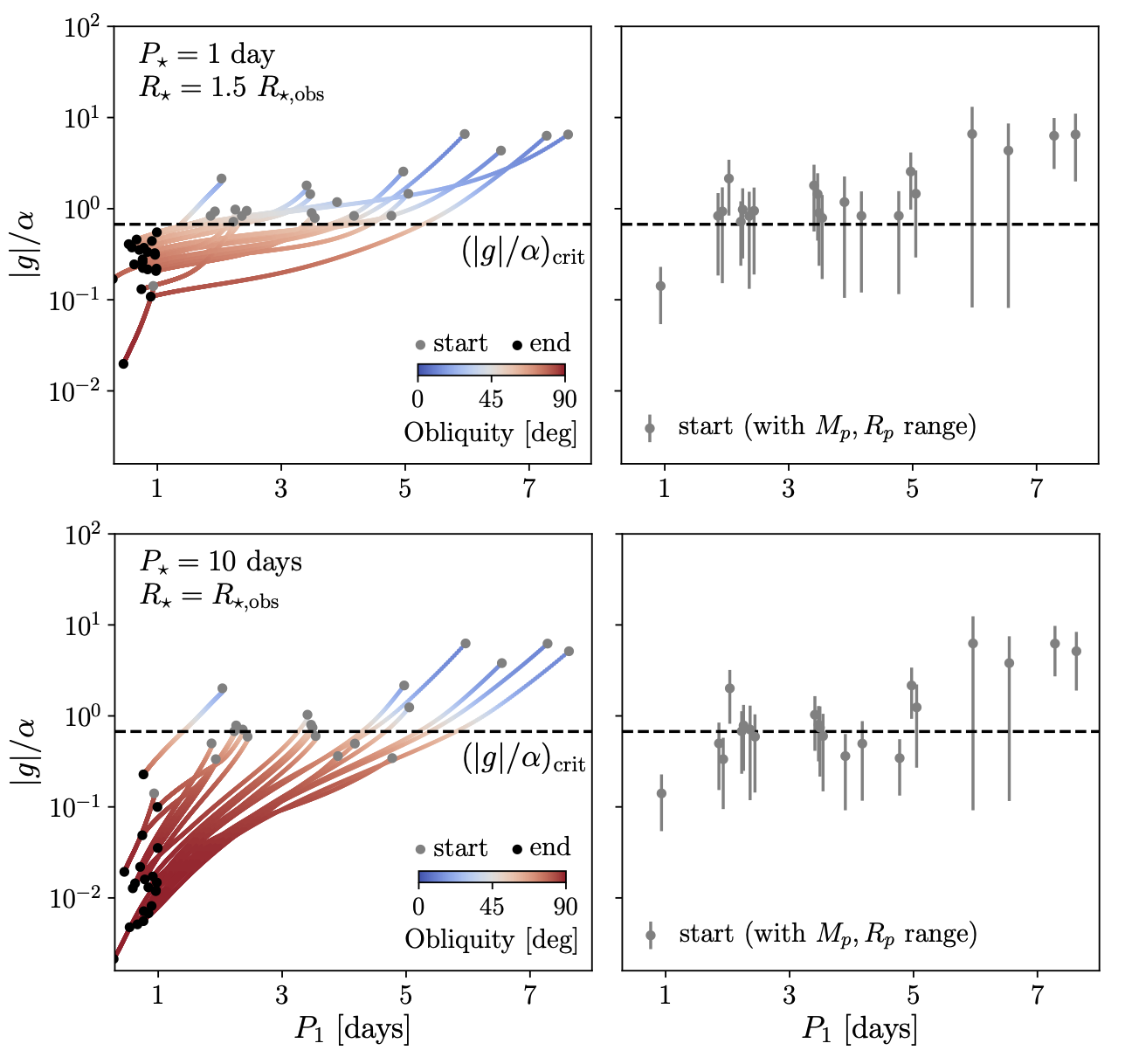}
\caption{Cassini state theory in the context of the observed USPs in multi-transiting systems. We restrict the depicted sample to systems with $P_2<10$ days, leaving out five systems. Each subplot shows $|g|/{\alpha}$ vs. $P_1$, with the top/bottom rows corresponding to stronger/weaker stellar quadrupoles. The left panels show the evolution of $|g|/\alpha$ across the full $P_1$ range, with the colorbar indicating the Cassini state 2 obliquity. The ``start'' ($P_{1,\mathrm{start}}=P_2(P_2/P_1)^{-1}_{\mathrm{min}}$) and ``end'' ($P_{1,\mathrm{end}}=P_{1,\mathrm{obs}}$) points are accentuated with gray and black dots. The right panels show just the the ``start'' points, where the errorbar represents the range obtained from varying the planetary masses and radii within $1\sigma$ of their median estimates. The starting $|g|/{\alpha}$ values are typically greater than $(|g|/{\alpha})_{\mathrm{crit}}$ (indicated by the horizontal dashed lines), while the smaller ending ratios are consistent with high obliquities, indicating that USPs would have already reached tidal breaking.} 
\label{fig: period_evolution_USP_multis}
\end{figure*}

While our theoretical analysis has shown that obliquity-driven tidal migration can apply to inner members of prototypical \textit{Kepler} multi-planet systems, we can gain further insight by examining observed USP systems in the context of Cassini state theory. We use the sample of USPs in multi-transiting systems from \cite{2018NewAR..83...37W}; these systems are depicted in their Figure 6. There are 29 systems in total: 23 from the \textit{Kepler} prime mission, 4 from the \textit{K2} mission, and 2 others.\footnote{We note that one system, K2-106, was duplicated in \cite{2018NewAR..83...37W}. In addition, we will leave off the 4.25 hr period KOI-1843.03 and WASP-47 e, as these are both outlier cases.} First we gather the orbital periods, radii, and stellar masses. For \textit{Kepler} systems, we use the parameter tables from \cite{2018AJ....156..264F} when available.
%who combined \textit{Gaia} DR2 parallaxes  \citep{2018A&A...616A...1G} and the California-\textit{Kepler} Survey \citep[CKS,][]{2017AJ....154..107P, 2017AJ....154..108J} to obtain more precise system parameters. 
Otherwise, we use the \textit{Kepler} Data Release 25 \citep{2018ApJS..235...38T}. For \textit{K2} systems, we use the tables from \cite{2020ApJS..247...28H}.
% who derived updated parameters using \textit{Gaia} DR2 and the Large Sky Area Multi-Object Fibre Spectroscopic Telescope (LAMOST) DR5 \citep{2012RAA....12.1197C}. 
Finally, for \textit{K2} systems not in this catalog and for all other systems, we use the NASA Exoplanet Archive (NEA, \citealt{2013PASP..125..989A}). 

In addition to periods, radii, and stellar masses, our calculation also requires planetary masses. Some but not all of the planets have mass constraints from radial velocities or transit timing variations. When available, we obtain these from the NEA. Otherwise, we use prediction tools to obtain mass estimates from the planetary radii. The USP masses are well-approximated using the Earth-like composition curve from \cite{2016ApJ...819..127Z}. (Recall that we used the same relationship in Section \ref{sec: System set-up and parameter space definition} to obtain $R_{p1}$ from $M_{p1}$.) Since the non-USP planets in the systems are not all rocky, we used the \texttt{Forecaster} code from \cite{2017ApJ...834...17C} to probabilistically estimate planetary masses from radii.

With all of the system parameters in hand, we can now examine the observed USPs in known multi-transiting systems in the context of our hypothesis. We begin by calculating $|g|/{\alpha}$ for observed USPs for both their present-day and theorized past orbits, assuming that they started with closer separations to their nearest companion planets. Following the approach in Section \ref{sec: Calculation of g/alpha using secular frequencies}, we obtain the secular eigenfrequencies and take $|g| = |g|_{\mathrm{max}}$. We do this for $P_1$ in the range $[P_{1,\mathrm{obs}}, P_2(P_2/P_1)^{-1}_{\mathrm{min}}]$, where $P_{1,\mathrm{obs}}$ represents the present-day period of the USP and $(P_2/P_1)_{\mathrm{min}}$ represents the minimum plausible initial period ratio between the proto-USP and its nearest neighbor. We take this parameter to be $(P_2/P_1)_{\mathrm{min}}=1.3$. The periods of the companion planets are held fixed at their present-day values. Next, we calculate the USP's spin-axis precession constant $\alpha$ (equation \ref{eq: alpha}) across the range in $P_1$. As before, the $\alpha$ calculation uses $k_2=0.4$ and $C=0.35$ as fiducial values; the result is not strongly sensitive to this choice. Finally, we calculate the ratio $|g|/{\alpha}$ across the $P_1$ range and obtain the Cassini state 2 obliquity according to equation \ref{eq: Cassini state relation}, assuming $I=10^{\circ}$ and $\omega = \omega_{\mathrm{eq}}$.

Figure \ref{fig: period_evolution_USP_multis} shows the evolution of $|g|/{\alpha}$ vs. $P_1$, with the obliquity indicated by the colorbar, for the set of observed USPs in multi-transiting systems. Since the stellar $J_{2\star}$ decreases over time and the migration time $\tau_{\mathrm{decay}}$ is also unknown, we illustrate the evolution using two different cases for the stellar quadrupolar moment. The top row uses values corresponding to early on in the system lifetime ($\lesssim 10$ Myr), when the host star is rapidly rotating and inflated \citep{2013ApJ...778..169B, 2014prpl.conf..433B}. We take $P_{\star} = 1$ day and $R_{\star} = 1.5 \ R_{\star, \mathrm{obs}}$ to represent fiducial values, where $R_{\star, \mathrm{obs}}$ is the present-day estimate. We note, however, that there is substantial uncertainty in these estimates; these values are simply to aid our order-of-magnitude calculations.  In contrast, the bottom row corresponds to later times ($\gtrsim 100-500$ Myr), where we use $P_{\star} = 10$ days and the present-day radius estimate, $R_{\star} = R_{\star,\mathrm{obs}}$. 

Broadly speaking, the results for these observed systems are consistent with our theoretical framework. At early times, the starting $|g|/{\alpha}$ (when the USP is close to its nearest neighbor) is typically greater than $(|g|/{\alpha})_{\mathrm{crit}}$, leading to inevitable capture into Cassini state 2, as described in Section \ref{sec: Cassini states and obliquity tides}. Moreover, at later times, the ending or present-day $|g|/{\alpha}$ of the observed USPs is in the range such that the planets' rapid, obliquity-driven tidal decay would have already stalled. That is, the present-day values of Cassini state 2 are beyond the tidal breaking limit identified in Section \ref{sec: Tidal breaking of Cassini states}, indicating that the USPs have already broken out of these states (assuming they were once in them), and the rapid orbital decay has ceased.

It is important to keep in mind that $P_{\star}$ and $R_{\star}$ evolve significantly during the first $\sim 10-100$ Myr of the star's lifetime \citep[e.g.][]{2014prpl.conf..433B}, such that examining $|g|/{\alpha}$ using fixed values for these quantities is only relevant in providing bounds on the dynamical evolution. For instance, the ending obliquities in the strong quadrupole case are generally smaller than the tidal breaking limit of Cassini state 2. This is consistent with the picture that the USPs reached their final orbits when $P_{\star}$ and $R_{\star}$ were evolving sometime after $\sim 10$ Myr.

\section{Limiting Factor: Angular Momentum Budget}
\label{sec: Limitation: angular momentum budget}
%(*Mention exterior giant, and separately multiple small close planets, eg Kepler-80. Rephrase. Make clear that it's 20 degrees in 2 planet case.*)

In the picture proposed thus far, we envision the inner planet to shrink its semi-major axis by a factor of $\sim2$ (Figure~\ref{fig: Secular_orbital_decay_examples}). The obliquity tides driving this orbital decay act by way of energy dissipation within the planet, but without the loss of angular momentum \citep{2007ApJ...665..754F}. For a circular orbit, the angular momentum normal to the plane depends only upon $a$, whereas secular interactions are unable to alter $a$. Thus, as the inner planet migrates, angular momentum conservation requires that the orbits become more aligned.\footnote{Angular momentum is also conserved in the case of a single planet with no additional perturbers by a non-zero initial obliquity. In that case, the planetary spin angular momentum is transferred to the orbital angular momentum as the planetary obliquity decays. Given the smallness of the planetary spin angular momentum relative to the orbit, an isolated planet can only migrate by a small amount from obliquity tides.} The end-state of perfect alignment places limits upon the extent of orbital decay. 

We recapitulate the arguments discussed in \citet{2007ApJ...665..754F}, first considering orbital angular momenta alone and next including stellar spin angular momentum. The total angular momentum $J$ of two planets on circular orbits is given by
\begin{align}\label{eq: Angular}
    J^2&=\big|\mathbf{L}_1+\mathbf{L}_2\big|^2
    \nonumber\\
    &=L_1^2+L_2^2+2L_1L_2\cos{I_{12}},
\end{align}
where $L_j=M_{pj}\sqrt{GM_\star a_j}$ is the angular momentum normal to the orbital plane of planet $j$ and $I_{12}$ is the mutual inclination between the two planets. Suppose that planet 1 decays from $L_{1,i}$ to $L_{1,f}$, while $L_2$ remains unchanged. The minimum attainable value of $L_{1,f}$ is found by setting the final mutual inclination to zero while imposing that $J^2$ is unchanged
\begin{equation}\label{eq: 2Planets}
\frac{(L_{1,f})_{\mathrm{min}}}{L_{1,i}}= \bigg[1+\bigg(\frac{L_2}{L_{1,i}}\bigg)^2+2\bigg(\frac{L_2}{L_{1,i}}\bigg)\cos{I_{12}}\bigg]^{\frac{1}{2}}-\frac{L_2}{L_{1,i}}.
\end{equation}
%Expanding in terms of masses and semi-major axes, this becomes 
%\begin{align}\label{eq: Pfinal}
%\bigg(\frac{a_{1,f}}{a_{1,i}}\bigg)^{\frac{1}{2}}&\geq \Bigg[1+\bigg(\frac{M_{p2}}{M_{p1}}\bigg)^2\bigg(\frac{P_2}{P_1}\bigg)^{\frac{2}{3}}\nonumber\\
%&+2\bigg(\frac{M_{p2}}{M_{p1}}\bigg)\bigg(\frac{P_2}{P_1}\bigg)^{\frac{1}{3}}\cos{I_{12}}\Bigg]^{\frac{1}{2}}\nonumber\\
%&-\bigg(\frac{M_{p2}}{M_{p1}}\bigg)\bigg(\frac{P_2}{P_1}\bigg)^{\frac{1}{3}}.
%\end{align}

For illustration, we assume that $L_{2}\gg L_{1}$. In this case, in order for the inner planet to migrate inwards from $a_{1,i}$ to $a_{1,f}$, the mutual inclination of the planets must satisfy $\cos{I_{12}}\lesssim L_{1,f}/L_{1,i}$. That is, the inner planet can reduce its orbit by a factor of two only if the initial mutual inclination is at least $45\,^\circ$. This is a rather stringent constraint upon the initial mutual inclination. It is a consequence of the fact that the direction of the inner planet's angular momentum vector must change in order to account for its decreasing magnitude. Note that the opposite extreme, when $L_2\ll L_1$, is even worse, with $a_{1,f}/a_{1,i}\rightarrow 1$, regardless of $I_{12}$.

Accordingly, in the case of two planets, angular momentum conservation demands large mutual inclinations if a USP is to result. However, the above picture has ignored the substantial angular momentum held within the stellar rotation. The stellar spin angular momentum, scaled by that of the inner planet, is given by

\begin{equation}
\begin{split}
\frac{I_\star M_\star R_\star^2 \omega_\star}{M_{p1}\sqrt{GM_\star a_1}}&\approx 30\bigg(\frac{M_\star}{M_\odot}\bigg)^{\frac{1}{2}}\bigg(\frac{R_\star}{R_\odot}\bigg)^2\bigg(\frac{P_\star}{10\ \textrm{day}}\bigg)^{-1}\\
    &\times\bigg(\frac{a_1}{0.1 \ \mathrm{AU}}\bigg)^{-\frac{1}{2}}\bigg(\frac{M_{p1}}{5\ M_\oplus}\bigg)^{-1}\gg1.
\end{split}
\end{equation}
The star therefore possesses significantly more angular momentum than typical close-in super-Earths. Stellar hosts of USPs spin-down over time, modifying the ratio above by a factor of $\sim 2$ as the system evolves over 0.1-1\,Gyr timescales \citep{2014prpl.conf..433B}. Even so, the star remains the dominant angular momentum source.

With the inclusion of the stellar angular momentum, we return to the problem of conserving full-system angular momentum. In this scenario, $J$ comprises three individual angular momenta: $L_1$ and $L_2$ as before and the stellar spin angular momentum, $L_\star$. Instead of following all three vectors, we sum the planetary orbital angular momenta into a single vector $\mathbf{L}_p\equiv \mathbf{L}_1+\mathbf{L}_2$, which is easily generalized to $N$ planets. 

Thus, we repeat the calculation above, conserving the angular momentum supplied by $L_p$ and $L_\star$ in an analogous manner to $L_1$ and $L_2$. We assume that $L_\star$ does not change in the process (again, ignoring stellar spin-down). Conserving angular momentum during realignment yields the following constraint 
\begin{equation}\label{eq: Pfinal}
\frac{L_{p,f}}{L_{p,i}}= \bigg[1+\bigg(\frac{L_\star}{L_{p,i}}\bigg)^2+2\bigg(\frac{L_\star}{L_{p,i}}\bigg)\cos{I_{p\star}}\bigg]^{\frac{1}{2}}-\frac{L_\star}{L_{p,i}},
\end{equation}
which is analogous to equation~\ref{eq: 2Planets}, with $I_{p\star}$ being the stellar obliquity. Importantly, the ratio $L_{p,f}/L_{p,i}$ no longer depends only upon the inner planet, but upon \textit{all} of the planets. If we take the limit of large stellar angular momentum ($L_\star\gg L_p$), we find that the minimum initial stellar obliquity in the two-planet case is given by
\begin{equation}\label{eq: Icrit}
\cos{I_{p\star}}<\frac{L_{p,f}}{L_{p,i}}=\frac{M_{p1}\sqrt{a_{1,f}}+M_{p2}\sqrt{a_2}}{M_{p1}\sqrt{a_{1,i}}+M_{p2}\sqrt{a_2}}.
\end{equation}
This is less stringent than the required mutual inclination determined when only considering the planetary angular momenta. The addition of stellar angular momentum thus allows the inner planet to migrate further inwards for the same amount of inclination. However, the important inclination $I_{p\star}$ now refers to the stellar obliquity, not to the planet-planet mutual inclination. 

\begin{figure}
\epsscale{1.1}
\plotone{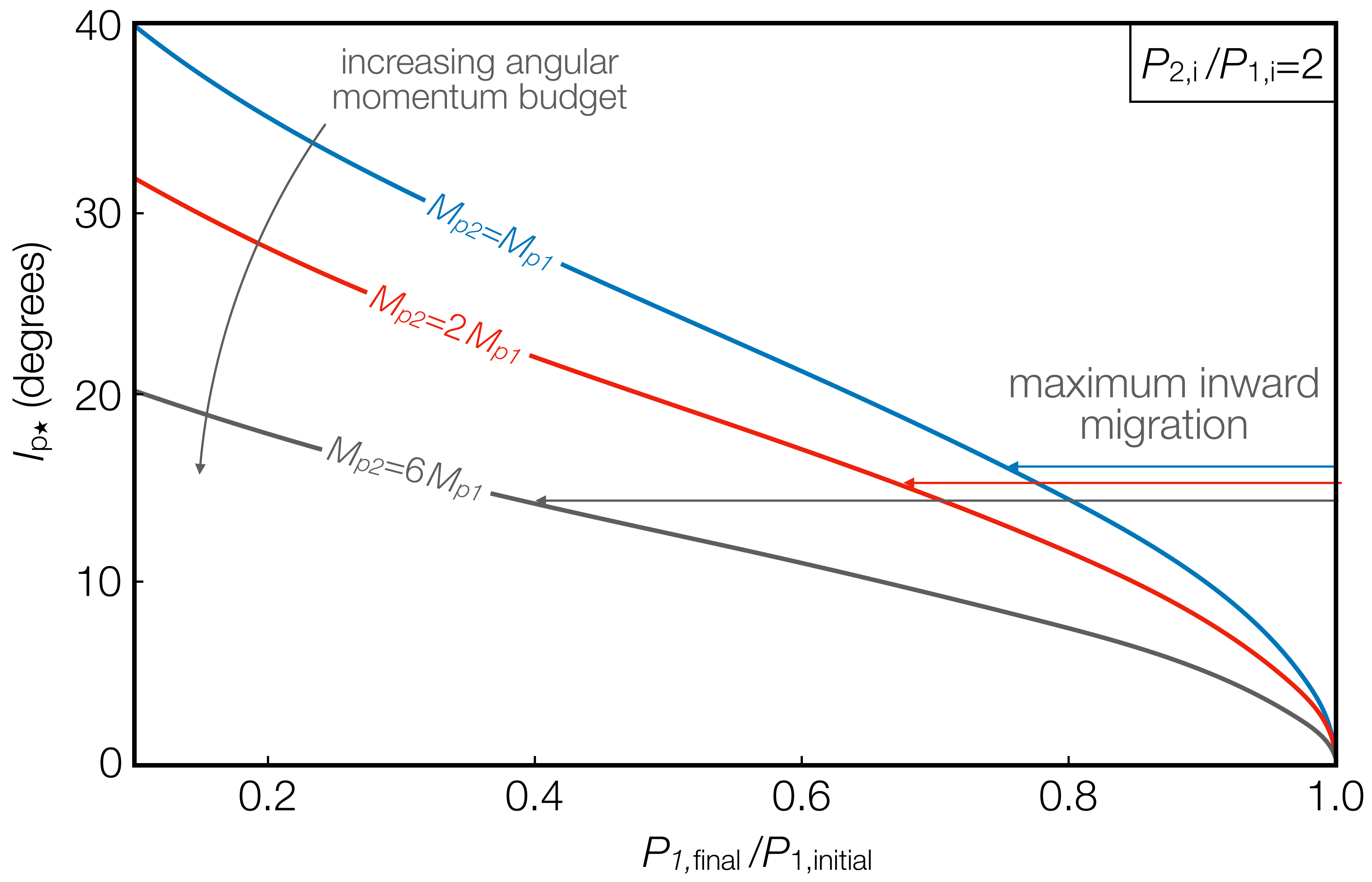}
\caption{The stellar obliquity in a two-planet system required in order to conserve angular momentum as a function of the maximum inward extent of migration for the USP. Inward migration is indicated by the final period divided by the initial period, and three cases are considered: $M_{p2}=M_{p1}$, $M_{p2}=2M_{p1}$ and $M_{p2}=6M_{p1}$. In general, the greater the angular momentum of the outer planet, the smaller the required initial stellar obliquity for significant inward migration. } 
\label{fig: Constraints}
\end{figure}

We illustrate the critical stellar obliquity in Figure~\ref{fig: Constraints}. Specifically, we choose three different mass ratios $M_{p1}/M_{p2}$ and plot the initial stellar obliquity required as a function of the ratio of the final to the initial period of the inner planet, $P_{1,f}/P_{1,i}$. With more massive exterior planets, a smaller inclination is required for a given degree of migration. In particular, if the exterior planet is twice the mass of the USP (red line), then a stellar obliquity of $\sim 20^\circ$ is required to reduce the period by a factor of two. An exterior planet of six times the USP's mass only requires about $10^\circ$. Moreover, if multiple low-mass planets reside exterior to the USP, the inclination requirements are further relaxed. Note also that the two planets may retain misalignments with one another, even once their summed angular momenta align with the star. We return to this point later.

%Physically, if the perturber (star or planet) possesses larger angular momentum, it needs to reorient less in response to receiving the same fraction of the inner planet's orbital angular momentum. 

In the discussion of angular momentum, we have ignored the influence of an exterior giant planet \citep{zhu2018super,bryan2019excess}, which would provide an even greater angular momentum sink than the host star, and substantially relax the inclination constraints. A distant giant would also modulate the eigenfrequencies $g$ to a degree comparable to, or less than, the host star's quadrupolar potential \citep{2020AJ....160..105S}. Cumulatively, angular momentum constraints require small, but reasonable stellar obliquities of $\sim 20^\circ$, or less, in order for the inner planet to become a USP.

\subsection{Example orbital evolution}
\label{sec: Example orbital evolution}

\begin{figure*}[ht!]
\plotone{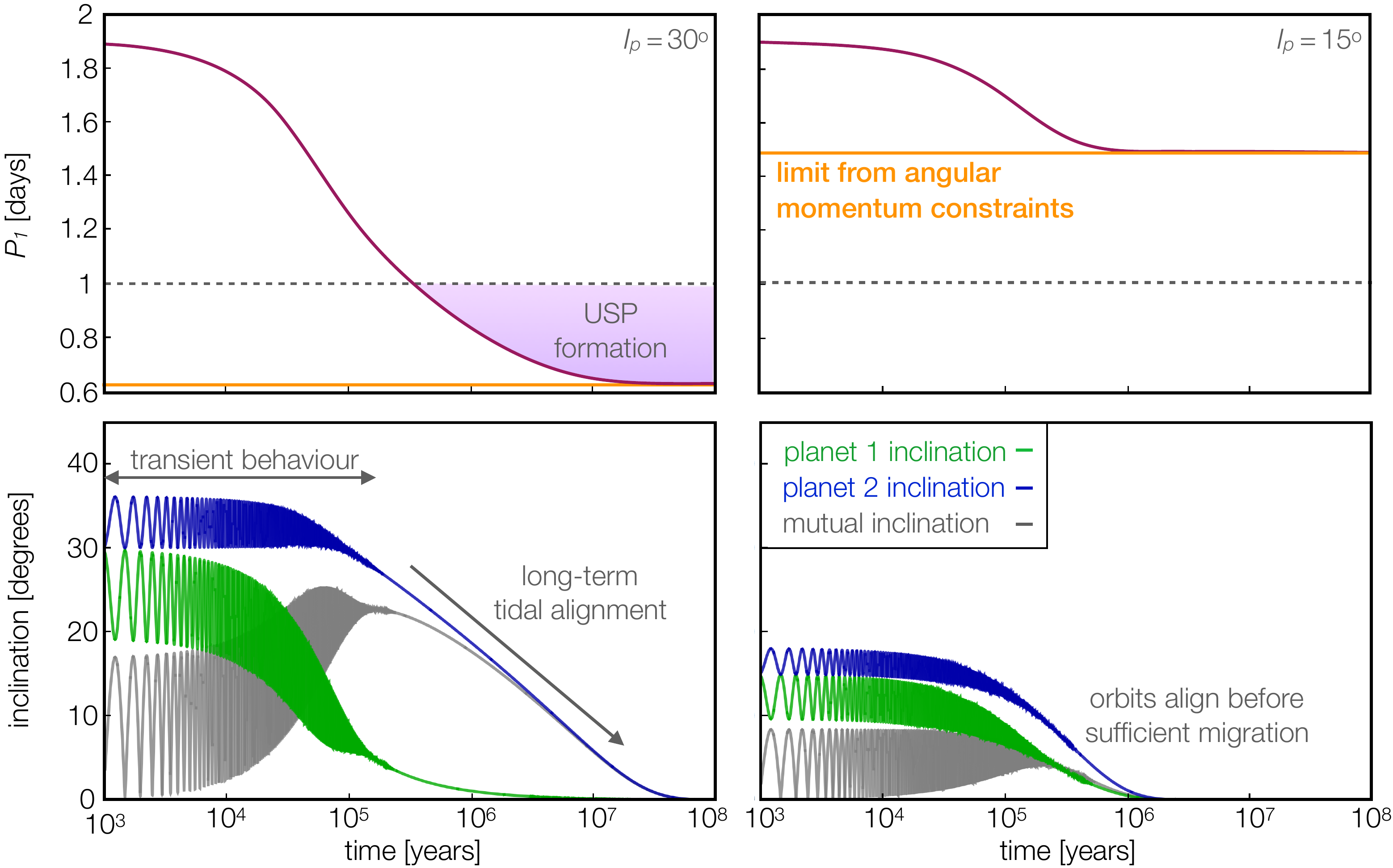}
\caption{Secular simulations of two-planet systems incorporating angular momentum-conserving tidal evolution. Each planet possesses $M_{p1}=M_{p2}=5\,M_\oplus$, and they are initialized with $a_1=0.03\,$AU and $a_2=0.05\,$AU. Tides are prescribed via equation~\ref{conserve}, such that the semi-major axis and inclination of the inner planet decay while conserving total angular momentum. On the left (right), both planets are initialized with inclinations of $30^\circ$ ($15^\circ$). We plot the inner planet's period in the top panels (magenta line) along with the expected minimum period reachable if all of the angular momentum deficit is exhausted (orange line; equation~\ref{eq: Pfinal}). The lower panels track the inclinations of the outer (blue) and inner (green) planets. Their mutual inclination is given in gray. Only the larger inclination case generates a USP (shaded region in the upper left panel). The inclination evolution begins displaying transient, oscillatory behaviour that is rapidly damped, followed by a longer-timescale decay of the outer planet's inclination. For the smaller inclinations on the right, the system is tidally aligned before the inner planet migrates enough to be classed as a USP.} 
\label{fig: secular}
\end{figure*}

To better understand the limitations imposed by angular momentum conservation, it is instructive to examine the inner planet's inward migration, while adhering to the above framework. Angular momentum is conserved by way of tidal torques upon the planet being transferred to the orbit. In order to account for these torques within a secular model, we present in Appendix \ref{sec: Secular model with tides} an extension of the secular model in Section \ref{sec: Calculation of g/alpha using secular frequencies} that incorporates angular momentum conservation into the inner planet's orbital decay (see also \citealt{chyba1989tidal}). This is accomplished by way of a forced decay of the orbital inclination over a timescale $\tau_I$ while the semi-major axis decays.

In this section, we solve the inclination and semi-major axis evolution equations~\ref{eq: Full} \&~\ref{eq: adot} subject to initial conditions~\ref{eq: initial} and $a_{1,i}$. We choose an inclination-damping timescale $\tau_I$ such that the inner planet becomes a USP (i.e. reaches $P_1\lesssim 1\,$day) within $\sim100$\,Myr. As discussed above, the true time taken to form a USP via obliquity tides can be longer. However, for the purposes of this simulation, the choice of $\tau_I$ is arbitrary provided that it exceeds the timescale of the secular oscillations. 

As illustrated in Figure~\ref{fig: tau_a_heatmap}, the tidal migration rate increases as the semi-major axis decays, leading to a runaway migration. Accordingly, we modulate the timescale of inclination-damping by a factor of $(a_1/a_{1,i})^5$, prescribing $\tau_I=0.1~\textrm{Myr}~(a_1/a_{1,i})^5$. We carry out two secular integrations, each lasting 100\,Myr for a star with $M_\star=M_{\odot}$, $R_\star=R_{\odot}$, and $J_{2\star}=10^{-4}$. Each planet is given the same mass of $M_{p1}=M_{p2}=5\ M_\oplus$, and they begin at semi-major axes $a_{1,i}=0.03\,$AU and $a_{2,i}=0.05\,$AU. From angular momentum conservation (see equation~\ref{eq: Icrit}), the minimum required tilt between the orbital and stellar angular momenta is $I_{p\star}\gtrsim 24^\circ$. Therefore, we illustrate the importance of the angular momentum constraint by choosing one case to begin above the critical misalignment, at $I_{p\star}= 30^\circ$, and the second case to possess insufficient inclination, with $I_{p\star}=15^\circ$. These tilts predict, respectively, innermost achievable periods of $P_{1,f}=0.63\,$days and $P_{1,f}=1.5\,$days. 

The results of our secular integrations are displayed in Figure~\ref{fig: secular}. In the top panel, we show the evolution of the inner planet's orbital period during tidal evolution. The horizontal line indicates the minimum period achievable due to angular momentum constraints (equation~\ref{eq: Pfinal}). The bottom panels track the inner (green) and outer (blue) planetary orbital inclinations, with their mutual inclinations shown in gray. Left panels correspond to the greater initial stellar obliquity of $I_{p\star}=30^\circ$ and the right panels show $I_{p\star}=15^\circ$.

As expected from the discussion above, only when the initial stellar obliquity is sufficiently large is the inner planet able to migrate far enough inwards to become a USP, with $P_1 < 1$\,day (left). From the bottom left panel, we see that the evolution is characterized by transient, oscillatory behaviour for $t\lesssim 0.3\,$Myr, or a few inclination decay timescales. During this transient period, the system's evolution is governed by a mix of two eigenmodes. At this stage, the inner planet's inclination decays faster than the outer planet's and the mutual planet-planet inclination grows. 

Physically, the early growth in mutual inclinations arises from the inner planet becoming more dominated by the stellar quadrupole during its inward migration \citep{2020ApJ...890L..31L,becker2020origin}. Its orbit reorients to the stellar equator, while the outer planet remains inclined. Thus, if the inner planet becomes a USP at this stage, the two planets will exhibit a large mutual inclination, as observed \citep{2018ApJ...864L..38D}. A USP formed in this way is predicted to exhibit a low misalignment with respect to the stellar spin-axis, lower than its exterior companion. We return to this prediction in Section~\ref{sec: Predictions}. 

After the initial transient evolution, the system collapses onto a single eigenmode \citep{zhang2013secular,2019MNRAS.488.3568P}, as the outer planet's inclination begins to undergo substantial decay. The inner planet's period falls well below $P_1=1$\,day while the mutual inclination between the planets remains high. In this simple example, tidal migration continues until $P_1=0.6$ days, aligning the orbits, but as discussed above the planet typically breaks out of the Cassini state before reaching the ultimate tidal end-state (see Figures~\ref{fig: Breaking_obliquity} and \ref{fig: Secular_orbital_decay_examples}). 

When insufficient orbital inclination exists (right panel of Figure~\ref{fig: secular}), the orbits tidally realign before the inner planet migrates below $P_1=1\,$day. Once this state is reached, the planetary obliquity falls to zero, and tidal migration stalls. Accordingly, angular momentum places a strong constraint upon USP formation via obliquity tides. However, here we considered constraints due to one exterior planet at $0.05\,$AU. Angular momentum constraints become significantly less restrictive as the outer planet's angular momentum increases (Figure~\ref{fig: Constraints}) or if the number of exterior planets increases. Moreover, if some fraction of the orbital decay occurs through eccentricity tides \citep{2019AJ....157..180P, 2019MNRAS.488.3568P}, this will also weaken the constraint, since in that case orbital circularization would contribute to angular momentum conservation.

%(*Maybe we should mention the Dai et al. and Li et al. mutual inclination results here? Or come back to it in the discussion? We need to think about whether USPs should be preferentially formed in systems with large mutual inclinations?*)

\section{Discussion}
\label{sec: Discussion}

In this work, we have studied the production of USP planets via a new theoretical mechanism called \textit{obliquity-driven tidal migration}. We have shown that the mechanism can operate upon the innermost planets of prototypical \textit{Kepler} multi-planet systems, turning them into USPs via runaway orbital decay triggered by a forced Cassini state obliquity. In order to present a coherent picture, however, our analysis utilized simplifying assumptions that should be expanded upon with future study. Our idealized scenario has not tackled spin dynamical chaos and all aspects of the system's early evolution. We will briefly discuss these areas in Sections \ref{sec: Obliquity chaos} and \ref{sec: Early system evolution} below. Moreover, dissipation from obliquity tides can occur simultaneously with eccentricity-driven tidal dissipation. While both sources may play an important role in USP production, our study has isolated the effects of obliquity tides. The prospects of comparing these mechanisms may improve with further observations; we offer some predictions in Section \ref{sec: Predictions}.

\subsection{Obliquity chaos}
\label{sec: Obliquity chaos}
The solution for a multiple-planet system's inclination/node evolution generally involves a set of frequencies $\{g_i\}$ and amplitudes $\{I_{ji}\}$, such that each planet's orbital evolution can be approximated by a superposition of these modes. As discussed several times throughout this work, a planet's Cassini state may be established with any one of these frequency modes, typically that which is closest to the planet's spin-axis precession constant $\alpha$. However, the process of identifying the dominant frequency mode is complicated \citep[e.g.][]{1974AJ.....79..722P}, and we have not addressed it in this work. 

Moreover, when $\alpha$ is close to several orbital eigenfrequencies, or combinations of these frequencies, obliquity chaos may result from secular spin-orbit resonance overlap \citep{2019A&A...623A...4S}. All of the terrestrial planets in the Solar System have wide ranges of possible spin states in which their obliquities undergo large-amplitude chaotic variations \citep{1993Natur.361..608L}. Although Mars is the only planet that is currently undergoing strong ($\sim60^{\circ}$ amplitude) chaotic variations \citep{1973Sci...181..260W, 1993Sci...259.1294T}, the obliquities of the other terrestrial planets were likely chaotic in the past as well. The obliquities of Mercury and Venus have been stabilized by tides \citep[e.g.][]{1974AJ.....79..722P, 2003Icar..163....1C}, whereas Earth's obliquity would undergo chaotic variations if not for the Moon's stabilizing influence \citep{1997A&A...318..975N, 2014ApJ...790...69L}.

More work must be done to understand the prevalence of chaos in obliquity dynamics of short-period exoplanets, as well as the stability of Cassini states when multiple close frequency modes are present. Such dynamics could affect our proposed USP production mechanism. For instance, chaotic dynamics could prematurely knock planets out of their high-obliquity Cassini states before the planets have fully migrated. Chaos is not necessarily destructive for the mechanism though, since high-amplitude obliquity variations -- such as those experienced by Mars -- would still lead to large-scale, obliquity-driven orbital decay. Future studies of both chaos and tides within short-period planets will help inform these potential scenarios.

\subsection{Early system evolution}
\label{sec: Early system evolution}

If USP production happens early ($\lesssim 1$ Gyr), as we have postulated, multiple system parameters may be changing simultaneously during the orbital migration process. The resulting evolution is more complex than we have depicted. For instance, the star's initially rapid rotation starts slowing early on. The large initial $J_{2\star}$ aids the inward migration, since it yields faster $g$ frequencies, such that the orbital decay can go further before Cassini state 2 breaks (Figure \ref{fig: period_evolution_USP_multis}). In this work we did not parameterize stellar spin down, but instead considered extremes.

In addition, if the planet is born with a primordial H/He envelope, it will undergo early mass loss through thermal mechanisms \citep[e.g.][]{2017ApJ...847...29O, 2018MNRAS.476..759G}. This mass loss would only be a few-\% effect in $M_{p1}$ (thus not affecting the $g$ or $\alpha$ frequencies much), but the associated change in $R_{p1}$ could be a factor of two or more. For fixed $a_1$, a decrease in $R_{p1}$ would shrink $\alpha$ and the resulting Cassini state 2 obliquity. A shrinking $R_{p1}$ during migration could make the orbital decay go further, since it would delay the transition from Cassini state 2 to state 1. At the same time, however, the mass loss could change $Q$ and $k_2$ unpredictably, as could the strong interior heating (which might lead to a partially molten planet). These effects would be interesting to expand upon in a future exploration.

\subsection{Observational comparisons \& predictions}
\label{sec: Predictions}
Both obliquity tides and eccentricity tides are important components of the overall tidal dissipation rate \citep[e.g.][]{2010A&A...516A..64L}. There is no reason why they can't operate simultaneously within the same system or with one dominating over the other in specific systems. However, in terms of distinguishing the obliquity tides mechanism of USP production from the corresponding eccentricity-based mechanisms \citep[e.g.][]{2019AJ....157..180P, 2019MNRAS.488.3568P}, we can highlight several observational predictions of our hypothesis. 

First, the obliquity tides mechanism predicts that USPs should frequently have planetary companions with $P < 10$ days, whether or not they are co-transiting. This can be relaxed when the companions are more massive, so the condition is better stated as: if the USPs initially started with typical separations from their nearest neighbors, they should have had $|g|/{\alpha} \gtrsim 1$. Our theory predicts no clear trends with present-day eccentricities, whereas the eccentricity-based mechanisms may expect remnant eccentricity enhancements for the companion planets to USPs. This would be interesting to check with observational constraints of eccentricities with radial velocities, transit timing variations \citep[e.g.][]{2017AJ....154....5H}, or stability analyses \citep{2020arXiv200706521T}. 

Whereas our theory does not require large eccentricities, we do require non-zero inclinations. More specifically, in order for the inner planet to migrate inwards while conserving angular momentum, it must transfer some of its angular momentum to other planets or the star. Typically, the star constitutes a sufficiently large sink of angular momentum when its stellar obliquity exceeds $\sim20^\circ$ in the two-planet case. However, the presence of additional exterior small planets relaxes this constraint, as would the inclusion of a massive distant planet. As the USP migrates inwards, it aligns with the stellar spin-axis, and it does so faster than its exterior planets. Consequently, we suggest that USPs will tend to be misaligned with their exterior planetary companions, but aligned with the stellar spin axis. The former of these features is observed \citep{2018ApJ...864L..38D}. The latter could be tested through stellar obliquity measurements, with the larger, misaligned USP companions being more promising observational targets.

%(*Comment on Dai et al. inclination trends in the context of this theory. Should USPs be preferentially produced in systems with large mutual inclinations?*(*Note: The USP should have a low stellar obliquity whereas the outer planet shouldn't, cite Li and Dai *)Require some intiial mialignment. Not only do Dai come from J2, but also from the requirement that the mutual. Also, the further in the planet goes, the more aligned to 2 planets are. This may be inconsistent with Dai!)

An important aspect of our theory is that USP migration is stalled when the planet breaks out of Cassini state 2 at high obliquity. Given that migration happens rapidly relative to the total system lifetime, we predict that the majority of currently-observed USPs are not undergoing significant orbital decay. This is unlike the prediction for a current high planetary obliquity of hot Jupiter WASP-12 b \citep{2018ApJ...869L..15M}, which was hypothesized to perhaps be experiencing ongoing tidal decay from obliquity tides.
%That is, they should have already experienced the tidal breaking mechanism leading to destabilization of the high-obliquity Cassini state, and they should be settled in low-obliquity states. 
An additional consequence of tidal breaking is that this mechanism preferentially produces USPs closer to $P \sim 1$ day, as opposed to even shorter periods. Observationally, \cite{2014ApJ...787...47S} find that the USP occurrence rate peaks at 1 day and decreases with the orbital period, in agreement with our model. 
 
As a final observational connection, we recall that the obliquity tides framework predicts that USPs should be more common around smaller mass stars (Section \ref{sec: Identifying the susceptible parameter space for USP production}). This is consistent with the empirical trend among USP stellar hosts \citep{2014ApJ...787...47S}.  

\section{Conclusion}
\label{sec: Conclusion}

%The ultra-short period planets reside in some of the most extreme environments of any known exoplanets. Their origins are mysterious, as it's unlikely that they could have formed on their current orbits. Observations have provided a number of insightful clues. USPs are consistent with having originated as the innermost members of typical short-period, tightly-packed, multiple-planet systems. However, USPs have significantly larger period ratios with respect to their nearest neighbors than the typical period ratios of tightly-packed systems. Accordingly, a likely scenario is that USPs formed with typical separations at the inner edge of such systems, before becoming separated off and moving inwards. Such a process could happen readily through orbital decay due to star-planet tidal interactions. Thus, in our framework, USPs are the planets that acquired a fast enough initial migration to experience rapid runaway decay. (*That last sentence could be omitted.*)

The ultra-short period planets reside in some of the most extreme environments of any known exoplanets. They are unlikely to have formed in their current orbits, but observations have provided a number of clues as to their true origins. Namely, USPs are often found as the innermost members of otherwise typical short-period, tightly-packed, multiple-planet systems. However, USPs have anomalously large period ratios with respect to their nearest neighbors. Accordingly, a likely scenario is that USPs formed with typical separations at the inner edge of such systems, before becoming separated off and moving inwards. Such a process could happen readily through orbital decay due to star-planet tidal interactions. 

In this work, we introduce \textit{obliquity-driven tidal migration} as a robust mechanism for producing this orbital decay. The crucial idea is that, when planetary orbits are mutually-inclined, the equilibrium values of the obliquities are non-zero, often significantly so. The framework of our theory may be summarized as three key steps (Section \ref{sec: Cassini states and obliquity tides}). 
%This is unlike the orbital eccentricity, for which the equilibrium is generally $e=0$. 
First, tidal dissipation quickly forces the planetary spin vectors to assume their equilibrium configurations, which are called Cassini states. Specifically, when the planet is locked in Cassini state 2, the obliquity can be significantly enhanced. Second, the ensuing inward tidal migration forces an even larger obliquity, resulting in a runaway orbital decay process. Third and finally, the runaway is halted when the high obliquity state is tidally destabilized, and the obliquity inevitably damps down to a lower state. The orbital migration thus stalls, leaving USPs at their present-day close-in orbits.

We presented a secular analysis of close-in, multi-planet systems and outlined the region of parameter space in which the innermost planet is most susceptible to becoming a USP. The mechanism occurs most readily when the proto-USP's initial semi-major axis is $a_{1,i} \lesssim 0.04-0.05$ AU. We showed that USP production is more efficient around smaller mass stars (Section \ref{sec: Identifying the susceptible parameter space for USP production}), in the same direction as the observational trend for smaller stars to more frequently host USPs. Our primary analysis was focused on multi-planet systems with similar masses, but this mechanism can apply to a range of configurations, particularly systems where the exterior planets are more massive than the proto-USPs. Such systems make the theory somewhat more flexible (Section \ref{sec: Unequal planet masses}) and less confined by the restraints of system angular momentum preservation (Section \ref{sec: Limitation: angular momentum budget}).

The observed USPs in multi-transiting systems present consistencies with our theoretical framework (Section \ref{sec: Observed USP Planets in Multi-transiting Systems}). By tracking the orbits of these USPs back to smaller initial separations with their nearest neighbors, we find that many would have been forced to enter Cassini state 2, the high obliquity state that can lead to runaway orbital decay. There are, however, some subtle uncertainties on this conclusion based on the range of timescales for stellar spin-down.

Further observational studies of USP systems can help investigate the efficacy of this mechanism. USPs should be found with close companions with $P\lesssim10$ days, which would not be expected to exhibit any strong trends with eccentricities. In contrast, we predict that the USP is modestly inclined with its next nearest neighbor, but its orbit should reside closer to the stellar equatorial plane. Moreover, additional theoretical analyses, particularly a better understanding of the prevalence of chaos in the obliquity dynamics of short-period planets, will also help constrain or even falsify our hypothesis.

Broadly speaking, obliquities are fundamental properties of planetary bodies, driving the seasons and the tides alike. After centuries of studying the obliquities of planets and satellites within our Solar system, perhaps the ultra-short period planets are providing a signal to the surprising role of obliquities in sculpting the architectures of exoplanetary systems. 

\section{Acknowledgements}
We thank Josh Winn and Cristobal Petrovich for helpful comments on a draft of the manuscript, and Greg Laughlin, Daniel Jontof-Hutter, Fei Dai, and Yubo Su for inspiring questions that improved this work. We also thank the anonymous referee for their careful review. S.M. was supported by NASA through the NASA Hubble Fellowship grant \#HST-HF2-51465 awarded by the Space Telescope Science Institute, which is operated by the Association of Universities for Research in Astronomy, Inc., for NASA, under contract NAS5-26555. S.M. was also supported by the NSF Graduate Research Fellowship Program under Grant DGE-1122492. C.S. thanks the 51 Pegasi b Heising-Simons Foundation grant for their generous support. This research has made use of the NASA Exoplanet Archive, which is operated by the California Institute of Technology, under contract with the National Aeronautics and Space Administration under the Exoplanet Exploration Program.

\newpage
\appendix
\section{Secular model with tides}
\label{sec: Secular model with tides}

In the secular calculations of Sections \ref{sec: Calculation of g/alpha using secular frequencies} and \ref{sec: Identifying the susceptible parameter space for USP production}, we assumed that a single secular mode $g$ forces a non-zero obliquity in Cassini state 2, driving tidal dissipation. Therein, the forced obliquity depends upon the orbital inclination.  However, real planetary systems are permeated with multiple secular modes, $\{g_i\}$, with a corresponding set of amplitudes, $\{I_{ji}\}$. Only one of these modes can force a Cassini state at any given time. However, the angular momentum constraints outlined in Section \ref{sec: Limitation: angular momentum budget} ensure that the orbital inclinations, together with the mode amplitudes, cannot remain fixed with time \citep{2019MNRAS.488.3568P}.

Here, we extend our secular formalism to allow for tidal evolution of the secular modes. Throughout, we assume the small angle regime (Laplace-Lagrange theory, \citealt{1999ssd..book.....M}) to hold. Under this approximation, the evolution of the complex inclination vector $\xi_j\equiv I_j \exp(\imath\Omega_j)$ is given by the following ODE \citep{1999ssd..book.....M,2019MNRAS.488.3568P}
\begin{equation}\label{eq: Full}%This is the complex conjugate of the equation used in our other paper
\frac{d \dot{\bm\xi}}{dt}=\imath\mathbf{M}(t)\bm\xi+\imath \bm{\nu_\star}\beta_\star,
\end{equation}
where $\beta_\star$ is the stellar obliquity (tilted along the real axis) and $\mathbf{M}$ is a matrix whose elements are derived below. We will assume that the stellar spin axis is fixed at $\beta_\star=0$, which is appropriate when $L_\star\gg L_p$. 

As a result of obliquity tides, the matrix $\bm{M}$ possesses both non-dissipative (real) and dissipative (imaginary) parts. The non-dissipative part is equivalent to matrix $\mathbf{B}$ defined in equations~\ref{eq: B_jj}, and arises from purely conservative gravitational interactions between the planets and stellar quadrupole. 

Obliquity tides act to reduce the inner planet's semi-major axis by way of energy dissipation while angular momentum is conserved. For the present calculation, we do not self-consistently model the evolution of the spin-axis. To do so is complicated by that fact that the equilibrium obliquity of the Cassini state depends upon $g$, which is itself evolving throughout the migration. Instead, we mathematically enforce angular momentum conservation as $a_1$ decreases \citep{chyba1989tidal}. Specifically, the angular momentum in the $z$-direction of the inner planet is given by $L_{1,z}=M_{p1}\sqrt{GM_\star a_1}\cos{I_1}$. Tides will cause $a_1$ to shrink, with the orbit tilting in order to preserve the total angular momentum. By solving $\dot{L}_z=0$ we find that
\begin{equation}\label{conserve}
\frac{\dot{a}_1}{a_1}=2\dot{I_1}\tan{I_1}\approx2I_1^2\bigg[\frac{\dot{I}_1}{I_1}\bigg],
\end{equation}
where the second equality arises via the small angle approximation. 

The relationship above allows either $\dot{I}$ or $\dot{a}$ to be specified in the problem. In contrast, the full problem requires each to be derived separately from the stellar obliquity, which is solved self-consistently \citep{2007ApJ...665..754F}. We specify the inclination evolution to follow
\begin{equation}
\frac{\dot{I}_1}{I_1}=-\frac{1}{\tau_I},
\end{equation}
such that $a_1$ evolves according to
\begin{equation}\label{eq: adot}
\frac{\dot{a_1}}{a_1}=2\frac{\xi_1\xi_1^*}{\tau_I}.
\end{equation}
With this specification, the elements of the matrix $\mathbf{M}$ are given by
\begin{align}
M_{jj}&=-\nu_j-\sum_{\substack{k=1 \\ k \ne j}}^{N}B_{jk}+\imath \frac{1}{\tau_I}\nonumber\\
M_{jk}&=B_{jk}.
\end{align}
The frequencies $B_{jk}$ are the same as those in equation \ref{eq: B_jj}, and $\nu_j$ is defined as
\begin{align}
%B_{ij} &\equiv \frac{1}{4}\frac{M_{pj}}{M_\star}n_i \alpha_{ij}\bar{\alpha}_{ij}b_{3/2}^{(1)}(\alpha_{ij})\nonumber\\
\nu_j \equiv \frac{3}{2}n_j J_{2\star}\bigg(\frac{R_\star}{a_j}\bigg)^2.
\end{align}

The above equations constitute simultaneous ODEs for the time evolution of $\bm\xi$ and $a_1$ and require three initial conditions. As discussed in Section \ref{sec: Limitation: angular momentum budget}, the maximum inward migration is determined by the misalignment between the planetary and stellar angular momenta. In the small angle regime, most of the angular momentum lies along the $z$-axis, which is given by
\begin{equation}
L_{z}\equiv L_p\cos{I_{p\star}}\approx \big(L_1\cos{I_1}+L_2\cos{I_2}\big).
\end{equation}
Using equation~\ref{eq: Angular} and solving for the mutual inclination, we obtain, to lowest order in inclinations, 
\begin{equation}
I_{p\star}\approx \frac{L_1 I_1+L_2 I_2}{L_1+L_2},
\end{equation}
assuming both orbits have the same $\Omega$ initially. 

For the sake of simplicity, we assume that initially $I_1=I_2$. However, the solution may be considered as a sum of two eigenmodes, with frequencies given by 
\begin{equation}
2g_{\pm}=M_{11}+M_{22}\pm \sqrt{4 M_{12} M_{21} + (M_{11} - M_{22})^2}.
\end{equation}
The above form is cumbersome, but physical insight may be gained by considering the limit where $L_1\ll L_2$ and $\nu_2\ll \nu_1$. In this regime, the imaginary part of $g_\pm$ take the approximate forms
\begin{equation}
\lambda_1\approx \frac{1}{\tau_I}\,\,\,\,\,\,\,\,\lambda_2\approx \frac{1}{\tau_I}\frac{B_{12}^2}{(B_{12}+\nu_1)^2}\frac{L_1}{L_2}\ll\frac{1}{\tau_I}.
\end{equation}
Accordingly, the first mode damps much faster than the second \citep{zhang2013secular}. Under the same degree of approximation as above, the eigenvectors take the approximate form
\begin{align}
\mathbf{v}_1 = \begin{pmatrix}-\frac{B_{12}+\nu_1}{B_{21}}\\1\end{pmatrix}\,\,\,\,\,\,\,\,\,  
\mathbf{v}_2= \begin{pmatrix}\frac{B_{12}}{B_{12}+\nu_1}\\1\end{pmatrix}.
\end{align}
Suppose that mode 1 damps rapidly, leaving the system dominated by mode 2, i.e., $\bm\xi\approx A \mathbf{v}_2\exp(i g_2 t)$ such that the amplitude of $I_1\approx I_2(B_{12}/B_{12}+\nu_1)$. This state is equivalent to planet 2 dominating the angular momentum budget, with the inner planet lying upon the Laplace Plane between the outer planet's secular potential and the stellar quadrupole \citep{tremaine2009satellite}.

Given the two-timescale nature of orbital evolution, we begin the system close to eigenstate 2 in order to capture the longer-timescale evolution. Therefore, as initial conditions, we choose
\begin{align}\label{eq: initial}
\xi_{1,i}&=I_{2,i}\bigg(\frac{B_{12}}{B_{12}+\nu_1}\bigg)\bigg|_0\nonumber\\
\xi_{2,i}&=I_{2,i}.
\end{align} 

Of course, the approximations used above to derive the damping rates are not strictly applicable here, but the qualitative conclusion remains unchanged; one mode damps much faster than the other as the inner planet finds a quasi-steady inclination intermediate between that of the outer planet and the stellar spin axis. From there, the system relaxes more slowly to the eventual star-aligned case. In reality, however, the system likely will not reach the well-aligned end-case, as the planetary obliquity breaks out of the Cassini state before this.

\bibliographystyle{aasjournal}
\bibliography{main}

\end{document}